\documentclass{emulateapj}
\usepackage{lscape}
\usepackage{rotating}
\usepackage{longtable}

\def\sun{\ifmmode\odot\else$\odot$\fi}

\def\HII{\hbox{H\,{\sc ii}}}

\def\H2{\hbox{H$_{2}$}}

\shorttitle{The Local LIRG NGC\,5135}
\shortauthors{Bedregal et al.}

\begin{document}

\title{Near-IR Integral Field Spectroscopy study of the Star Formation and AGN of the LIRG NGC\,5135}


\author{Alejandro G. Bedregal\altaffilmark{1}, Luis Colina\altaffilmark{1}, Almudena Alonso-Herrero\altaffilmark{1}, Santiago Arribas\altaffilmark{1}}
\altaffiltext{1}{Instituto de Estructura de la Materia (IEM), CSIC, Serrano
  121, E-28006 Madrid, Spain. Contact email:{\it bedregal@astrax.fis.ucm.es}}
   
\begin{abstract}

We present a study of the central $\rm 2.3\,kpc$ of NGC\,5135, a nearby Luminous Infrared
Galaxy (LIRG) with an AGN and circumnuclear starburst.  Our main results are based on 
intermediate spectral resolution ($\rm \sim 3000$-$4000$) near infrared data taken with the SINFONI 
integral field spectrograph at the ESO VLT. The ionization of the 
different phases of the interstellar gas and the complex structures of the star formation have
been mapped. Individual regions of interest have been identified and studied in
detail.

For the first time in this galaxy, we have detected the presence of a high excitation
ionization cone centered on the AGN by using the [SiVI]${\rm \lambda1.96\,\mu m}$ line. 
So far, this structure is the largest reported in the literature for this coronal line, 
extending (in projection) as far as $\rm \sim 600\,pc$ from the galaxy nucleus. In a complex spatial distribution, a variety of mechanisms are driving the gas
ionization, including SNe remnant shocks, young stars and AGN
photoionization. The excitation of the molecular gas, however, is mainly
produced by X-rays and SNe remnant shocks. UV-mechanisms like fluorescence
represent a marginal overall contribution to this process, contrary to the expectations 
we might have for a galaxy with a recent and strong star formation. Our SNe rate estimations from [FeII]${\rm \lambda1.64\,\mu m}$ are in excellent agreement with 6\,cm radio emission predictions. Typical SNe rates between $0.01$-$\rm 0.04\,yr^{-1}$ were found for individual $\approx$ 200\,pc-scale
regions, with an overall SNe rate of $0.4$-$\rm 0.5\,yr^{-1}$. Even though NGC\,5135 has suffered a recent starburst ($6$-$\rm 7\,Myr$ ago),
the data strongly suggest the presence of a second, older stellar population
dominated by red giant/supergiant stars. However, simple stellar population
models cannot sharply discriminate between the different populations.

\end{abstract}


\section[]{Introduction}\label{s:intro}

Since their discovery (Kleinmann \& Low 1970, Rieke \& Low 1972), the importance of low-$z$
 Luminous ($10^{11}L_\odot\le L_{IR}\le 10^{12}L_\odot$, LIRG) and Ultraluminous
($L_{IR}\ge10^{12}L_\odot$, ULIRG) Infrared Galaxies has been widely recognized. (U)LIRGs 
are gas rich galaxies (Solomon et al.\ 1997), where
an active nucleus and powerful starbursts coexist and contribute to the energy output (e.g.\ Genzel et al.\ 1998, 
Spoon et al.\ 2007). While LIRGs appear to be mostly spirals (Arribas et al.\ 2004, Alonso-Herrero et al.\ 2006a),
ULIRGs are strongly interacting systems and mergers (e.g.\ Bushouse et al.\ 2002) evolving into intermediate-mass
ellipticals (e.g.\ Genzel et al.\ 2001, Dasyra et al.\ 2006). 
Local (U)LIRGs have been proposed as
possible counterparts of the submillimeter population observed at higher
redshifts (Blain et al.\ 2002 for a review, see also Chapman et al.\ 2003,
Frayer et al.\ 2003, Egami et al.\ 2004). Also, cosmological surveys with $Spitzer$
have shown that the majority of IR selected galaxies at $z\le1$ are in the LIRG
class, while LIRGs and ULIRGs make a significant contribution to the IR
galaxy population and to the star formation at $1<z<2$ and $z\ge2$, respectively (Egami et al.\ 2004; Le Floch et al.\ 2004, 2005; P\'erez-Gonz\'alez et al.\ 2005; Caputi et al.\ 2007).

Detailed investigations of the physical properties, stellar populations, AGN-starburst connection and
gas flows on these complex systems can only be obtained through integral field spectroscopy (IFS). Initial
studies of small samples of (U)LIRGs based on 4-meter class telescope optical IFS have already been obtained (e.g.\ Colina et al.\ 2005, Monreal-Ibero et al.\ 2006, Alonso-Herrero et al.\ 2009). 
To extend these studies to larger samples, and also to the near-IR, we have started a survey of
 low-$z$ (U)LIRGs using state-of-the-art IFS
like VLT/VIMOS (Arribas et al.\ 2008) and VLT/SINFONI (visual and near-IR, respectively). This survey
({\tt SIRIUS}: {\tt \underline{S}}urvey of luminous {\tt \underline{IR}} galaxies with {\tt \underline{I}}ntegral
field {\tt \underline{U}}nit{\tt \underline{S}}) will allow us to characterize the kpc-scale ionization and kinematics
of a representative sample of low-$z$ (U)LIRGs covering a wide luminosity range, several morphologies from spirals
to interacting and advanced mergers, as well as different classes of activity. This will also form a local reference for
future IFS studies of high-$z$ IR galaxies with instruments such as the {\it Near-IR Spectrograph} ({\it NIRSpec}) and 
{\it Mid-IR Instrument} ({\it MIRI}) on board of the {\it James Webb Space Telescope} (Gardner et al.\ 2006).

As part of our survey, we present our first results with SINFONI showing the power
of near-IR integral field spectroscopy by
studying the local LIRG NGC\,5135. This is an SBab galaxy at
$z=0.01396$ (from {\tt NED}\footnote{http://nedwww.ipac.caltech.edu/}, at $\rm \approx
58.7\,Mpc$ assuming $H_0=70\,\rm km\,s^{-1}\,Mpc^{-1}$) which belongs to a group
of seven galaxies (Kollatschny \& Fricke 1989). Its dual nature as a
starburst hosting an AGN and its almost face-on sky orientation make NGC\,5135
an ideal prototype-object for detailed studies of these hybrid systems.

The first attempts to study NGC\,5135 was made 25 years ago
by Huchra(1983), Huchra et al.\,(1983) and Phillips et al.\,(1983), who classified
this object as a Seyfert\,2. A year later, Thuan (1984) found that this
galaxy is also a bright UV source, with mixed Seyfert and
starburst spectral characteristics (see also Kinney et al.\,1993). Ulvestad \&
Wilson (1989) observed NGC\,5135 at 6 and 20\,cm using the VLA, as part of a
larger sample of nearby Seyfert galaxies. An important diffuse radio component
was measured in this morphologically asymmetric object, which is in turn
aligned with the extended H$\alpha$ emission (Haniff et al.\,1988, Garc\'ia
Barreto et al.\,1996). Because of the lack of an accurate optical nucleus, Ulvestad \&
Wilson chose the brightest radio source as the reference for the central region. 

Almost 10 years later, Gonz\'alez Delgado et al.\ (1998) published a study
based on $Hubble$ $Space$ $Telescope$ ($HST$ from now on) and ground-based UV/optical imaging and spectroscopy to study possible starburst-AGN connections. They found independent pieces of
evidence pointing towards the presence of a dusty, compact and powerful starburst in the
central region (inner $\rm 2^{\prime \prime}$) of NGC\,5135, which dominates
the UV spectrum. From spectral modeling and colors, these authors
estimated an age of between $3$ and $\rm 6\,Myr$ for the last starburst episode. They
also concluded that the warm interstellar gas is outflowing, most likely
because of the starbursts. In the optical, these authors detected a nuclear
bar for this galaxy, which had been previously found in the near infrared (hereafter NIR) by Mulchaey et
al.\,(1997).

Using {\it Chandra} X-ray observations, Levenson et al.\ (2004) revisited
NGC\,5135. Two strong central sources were detected and the (AGN) nucleus was
correctly identified for the first time, showing that its important internal
extinction had made it difficult to recognize in previous works. They
highlighted the importance of the stellar processes even in X-rays, showing
that almost all the soft emission ($0.4$-$\rm 1\,keV$) has a stellar origin,
even nearby the AGN. Only the hard emission ($4$-$\rm 10\,keV$) is dominated
by the active nucleus.

Alonso-Herrero et al.\ (2006a,b) and D\'iaz-Santos et al.\ (2008) studied this
object as part of a sample of $\sim 30$ local
LIRGs. With high resolution mid and NIR images (Gemini, $HST$) in different
bands, they identified the individual $\HII$ \normalsize knots which
are hidden at shorter wavelengths. In NGC\,5135, compact Pa$\alpha$ emission
($\sim 1$-$\rm 2\,kpc$) from star formation was detected as well as a strong,
nuclear warm dust component in the {\it N}-band. 

The NGC\,5135 SINFONI data allowed us to address some of the issues
described above, as well as to show the scientific potential of our SINFONI
data set. On the other hand, issues such as the general role of (U)LIRGs in
galaxy formation and evolution will be better addressed with our entire sample
in future papers of the series. We leave for the following paper of the series
(hereafter, Paper\,II) a detailed kinematical study and mass estimation for
this galaxy. 
\\

The paper is organized as follows: in Sec.\,\ref{s:obs} the observations
(\ref{ss:subs11}), data reduction (\ref{ss:subs12}) and analysis
(\ref{ss:subs13}) are described. Also, several maps are presented
(\ref{ss:subs14}) and specific
regions of study are identified (\ref{ss:subs15}). In Sec.\,\ref{s:s2}
the main results and discussion are shown, including a study of the central
gas and stellar structure (\ref{ss:subs23}), internal extinction structure (\ref{sss:subs232}), gas ionization mechanisms and young stellar
population age (\ref{ss:subs24}), further characterization of the stellar
population ages by using stellar absorption features (\ref{ss:subs26}), the H$_2$
excitation mechanisms (\ref{ss:subs25}) and estimations of the supernova rate
(\ref{ss:subs28}). Finally, in Sec.\,\ref{s:conclu} we present the conclusions.

\section[]{Observations and Data Reduction}\label{s:obs}

\subsection[]{Observing with SINFONI}\label{ss:subs11}
The observations were carried out in service mode with SINFONI (Eisenhauer et
al.\ 2003,  Bonnet et al.\ 2004), the NIR
integral field spectrometer mounted at VLT-Yepun (Cerro Paranal, Chile), with
no AO-assisted guiding. The galaxy was observed separately in the
{\it H} ($\rm 1.45-1.85 \mu m$ ) and {\it K} ($\rm 1.95-2.45 \mu m$) bands
during April and May 2006. Observing the {\it H} and {\it
  K}-bands independently provides a higher spectral resolution ($\sim$ 3000
and $\sim 4000$, respectively) compared to the combined {\it H}+{\it K} band
mode ($\rm \sim 1500$). The
full-width-at-half-maximum (FWHM from now on) as measured from sky lines is $\sim 6\,\rm \AA$
for the {\it H}-band and about $\sim 5\,\rm \AA$ for {\it K}. The dispersions are $\rm
2$ and $\rm 2.5\, \AA \, pixel^{-1}$ for {\it H} and {\it K}, respectively. The
 $\rm 64 \times 64$ spatial pixel (spaxels) field-of-view (FoV from now on) was
centered close to the region where the AGN is located (Levenson et al.\ 2004). With
a scale of $\rm 0.250^{\prime \prime} pixel^{-1}$, SINFONI provides a
two-dimensional field of $\rm 8^{\prime \prime} \times 8^{\prime \prime}$. 
 
For the {\it H}-band galaxy data, the total on-target integration time was 40
minutes (divided into 3 observation blocks, hereafter OBs), with an average
seeing (FWHM of the point-spread-function, hereafter PSF) and airmass of $\rm 0.54^{\prime \prime}$ and 1.24, respectively. For
the {\it K}-band data, the corresponding values were 25 minutes (2 OBs), $\rm
0.62^{\prime \prime}$ and 1.04 airmass. An entire {\it K}-band OB was
discarded because of the high seeing ($\rm >2^{\prime \prime}$). Given the
strong variations of the IR sky with time, the observing blocks were split into shorter on-source
integrations of 150 seconds each. Between them, we obtained identical
exposures of the sky background (ABBA pattern) by {\it nodding} the telescope
to a point shifted $\Delta \alpha = 42^{\prime \prime}$ and $\Delta \delta =
87^{\prime \prime} $ from the galaxy nucleus. In this way, each ``object
integration'' has a corresponding ``sky integration'' which was subtracted
during the pipeline processing to remove (to first order) the sky signal
from the final data cube. Also, the object and sky frames were {\it jittered}
within a box width of $\rm 1.5^{\prime \prime}$ to get rid of detector effects (bad pixels). 

\subsection[]{Data Reduction}\label{ss:subs12}

The basic data reduction was performed with the SINFONI Pipeline
(version\,1.8.2). Dark subtraction, flat-field normalisation, pixel linearity,
geometrical distortion and wavelength calibration corrections were applied for
each object-sky pair by using the products of SINFONI standard calibration
plan. The calibration products (e.\,g.\ darks and flats) are generated on a daily basis, with the
exception of the linearity and geometrical distortion data which are produced
monthly. The wavelength calibration was performed with Xe-Ar and Ne lamps for
the {\it H} and {\it K}-bands, respectively. The arc-frames include about 15
strong emission lines, well spread along the corresponding wavelength
ranges. We double-checked the wavelength calibrations on the individual
object-sky frames by tracing the position of about 10-15 strong sky-lines
along the detector. For the {\it H}-band data, we found a systematic offset in
wavelength of $\rm +1.6\pm 0.1 \AA$, independent of the data frame used for
the measurements. A similar situation affects the {\it K}-band data, where an
offset of $\rm +2.3\pm 0.3 \AA$ was found. Although these shifts are smaller
than the spectral resolution ($ \rm FWHM=5$-$\rm 6 \AA$) they correspond to almost 1\,pixel
offset. Therefore, we decided to correct for them given their amplitude and
systematic presence in the data. As a final check and after applying the offset
corrections, we compared for several spaxels the measured
and theoretical arc-line positions, repeating the measurements in different frames. From the distribution of their
differences, we estimated the ($\rm 3\,\sigma$) uncertainty in the wavelength
calibration to be $\le 1.3$ and $\rm \le 1\,\AA$ for the {\it H}- and {\it
  K}-band spectral ranges, respectively.

The background sky was subtracted for each individual $\rm 150\, s$ object
exposure. The SINFONI Pipeline does not only subtract the object/sky pairs but
it also applies other correction algorithms in order to minimise the sky line
residuals. Despite the fact that the sky line relative positions of each object/sky pair were
between spectral resolution limits, the sky frames were slightly shifted in
wavelength with respect to their object pairs to reduce the P-Cygni-type residuals.
In a second step, the pipeline includes an implementation of the sky
subtraction package described by Davies et al.\ (2007). This code allows to
separately scale different transitions of the telluric OH lines before sky
subtraction. After the correction, the sky-subtracted object frames present
very small sky-line residuals whose importance (in the worst cases) is purely cosmetic.

At this point, the signal (counts) from the object frames within an OB
(between 5 and 6 on-target frames per OB) were combined to generate the integrated
data-cubes for each band. Given the {\it jittering} applied during
observations, the pipeline re-aligned the individual object frame centers before combination. As a test
for these alignments, we generate a individual data-cubes from each object
frame; the position of the three more prominent peaks in the continuum (for
{\it H} and {\it K}-band) were traced by fitting two-dimensional elliptical
Gaussians and the offsets were re-calculated. The data-cubes generated with
the new offsets were in excellent agreement with the pipeline products. In
fact, the majority of the offsets used by the pipeline were recovered
by our method within a 10\% uncertainty. Consequently, we decided to stick to
the pipeline data-cube products, while our cube alignment method will be used
in following steps of the reduction.

As part of the calibration plan, spectrophotometric standard
stars were observed along with our target object in order to perform the flux
calibration. For each of the three {\it H}-band OBs an early-type star was
observed (Hip\,075902, Hip\,068496 and Hip\,065630 of spectral types B9V, B7II
and B3IV, respectively), while only one star was used for the two consecutive
{\it K}-band OBs (Hip\,078004, B3/B4V). The stellar spectra were extracted
within an aperture of $\rm 6\, \sigma$ of the best two-dimensional Gaussian fit of the light profile
(between $5$ and $6^{\prime \prime}$). The measured spectra were divided by
a theoretical spectrum for the corresponding stellar type, generating a
multiplicative ``sensitivity function'' to be applied to the object data. We use
the models of Pickles (1998) as our theoretical templates after scaling them
to the corresponding magnitudes listed in the Hipparcos catalogue (Perryman et
al.\ 1997).

 For the {\it H}-band stars, we found some differences between the three
 individual sensitivity functions. The corrections in the central range $\rm
 1.52$-$\rm 1.75\,\mu m$ are in very good agreement between the three
 functions, showing relative deviations within 5\%. In the extremes of the
 wavelength range ($\leq 1.50$ and $\rm \geq 1.75\,\mu m$), however, relative
 differences could be as large as 15\%. This suggests that the main source of
 uncertainty is not coming from Pickles' models (our ``real'' spectra) neither
 from peculiarities of the individual stars but from variations in the
 atmospheric transmission during observation (each star was observed just
 after one object OB, with airmass of  1.1, 1.3 and 1.5, respectively). We
 decided to apply the individual sensitivity functions to their corresponding
 OBs. Each of these functions has implicit the atmospheric transmission at a
 time and airmass close to that of the object observation. The results for the
 flux calibrated OB's data were very good, with no strong residuals from
 transmission variations. In any case, because the only {\it H}-band line we
 will use is  the [FeII]${\rm \lambda1.64\,\mu m}$ (essentially in the center
 of the band), the effects of the correction in the extremes of the wavelength
 range are purely cosmetic.

One problem we found with Pickles' models is that for many of our spectral
types the models do not include particular stellar features in the NIR
spectrum, but they just present a smooth continuum resembling that of a black body. If we
do not take care of this, the spectral features of the star will show up in
the galaxy spectrum. Fortunately, lines
from the Brackett series in absorption are essentially the only spectral
features we found in our stellar spectra. So they were modeled with Gaussians
and the stellar spectra were normalised before flux calibration. The residuals
from this process were negligible.

\begin{figure*}
\includegraphics[scale=0.35]{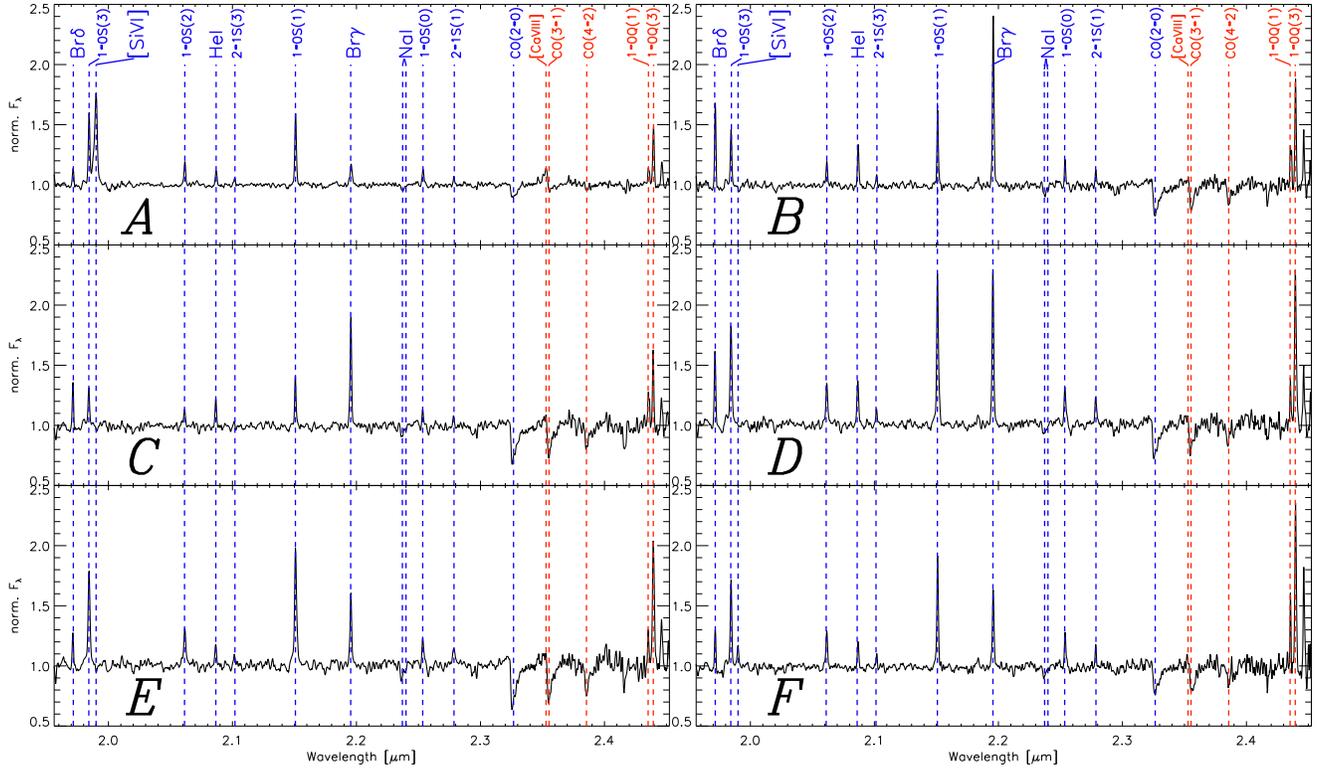}
\caption{\label{fig:specK}\small Observed-frame, normalized {\it K}-band
  spectra of six selected regions in NGC\,5135. The spectra were integrated
  within circular apertures of $\rm 0.62^{\prime \prime}$ diameter, corresponding to $\rm \approx 180\,pc$. Different emission lines and absorption
  bands are labeled. Blue labeled features are those used in this study.}
\end{figure*}

\begin{figure*}
\includegraphics[scale=0.35]{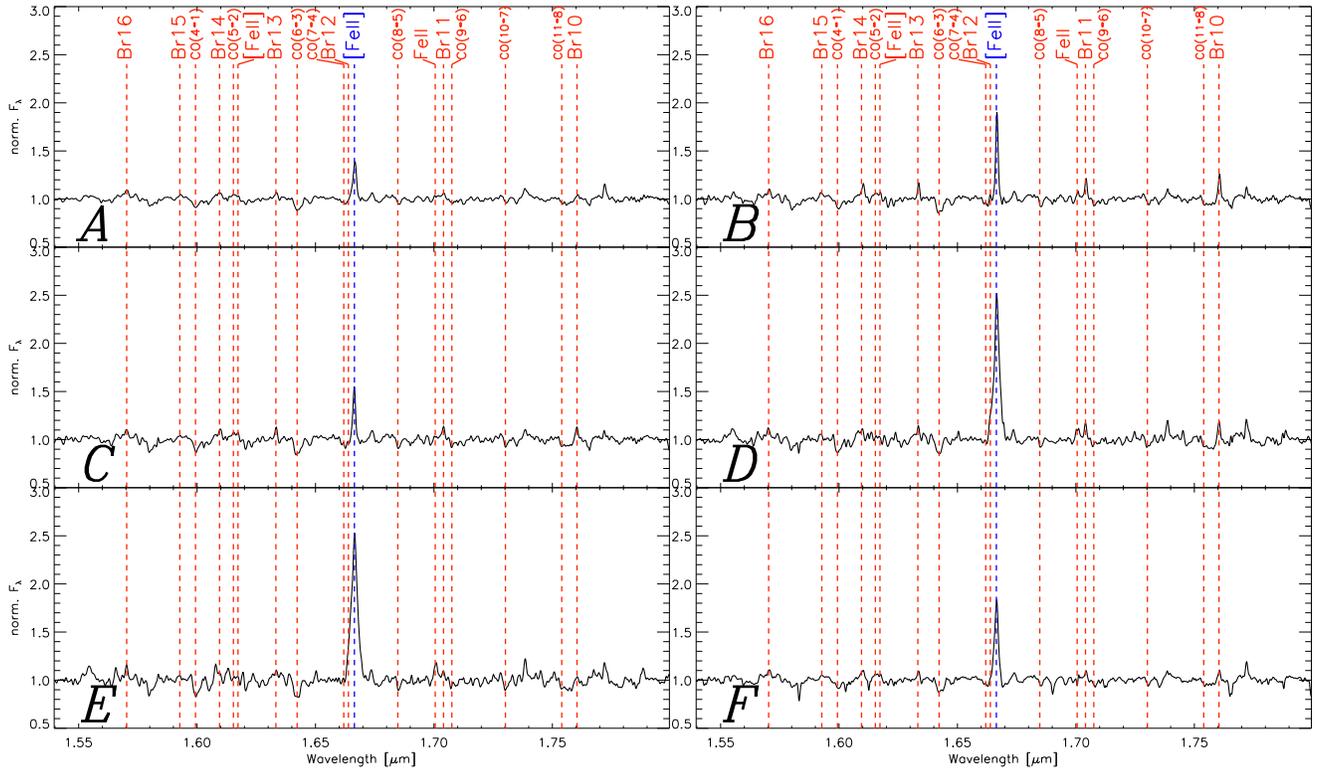}
\caption{\label{fig:specH}\small Observed-frame, normalized {\it H}-band
  spectra of six selected regions in NGC\,5135. The spectra were integrated
  within circular apertures of $\rm 0.54^{\prime \prime}$ diameter, corresponding to $\rm \approx 157\,pc$. Different emission lines and absorption
  bands are labeled. Blue labeled features are those used in this study.}
\end{figure*}

Once the OB's spectra were reduced and flux calibrated, we combined
the individual OBs in ``final'' data cubes for {\it H}- and {\it K}-bands. We
estimated the offsets between OBs by finding the three main peaks in the
continuum, as described above for the individual frame combination within an
OB. As a final check, we compared our integrated fluxes in {\it H} and {\it K}
with 2MASS data (Skrutskie et al.\ 2006). After adapting the spectral range to
the {\it $K_s$}-band used in 2MASS, selecting a galaxy's central region
comparable to our FoV and taking into account zero-point considerations, we
found a good agreement in fluxes within 10\% uncertainty for both bands. This
is consistent with our maximum error estimation (15\%) found between
individual OB's flux calibrations in {\it H}. In Fig.\,\ref{fig:specK} and \ref{fig:specH} we
present {\it K}- and {\it H}-band integrated spectra of six regions in
NGC\,5135 (see Fig.\,\ref{fig:maps} and Secs.\,\ref{ss:subs15} and \ref{ss:subs23} for further
details about the regions).

\subsection[]{Emission Line Fitting}\label{ss:subs13}

The (redshifted) line peak, flux, equivalent-width (EW) and FWHM were calculated from many of the emission lines
presented in {\it H}- and {\it K}-bands. Other parameters like
galactocentric velocities and velocity dispersions will be presented and
analysed in Paper\,II. 

A single Gaussian component provides a good fit for the individual emission
lines.  We have used a nonlinear least-squares minimisation algorithm
implemented in {\tt Image Data Language} (hereafter, {\tt IDL}) to perform the
fitting. This method is comparable to other standard algorithms used for line
fitting and implemented with other packages like {\tt IRAF}. For each spaxel,
local pseudocontinuum levels were calculated for each spectral feature. Two
bands were defined at each side of the emission lines and linear fits were
applied between their average flux values. In this way, our fitting algorithm
weights each pixel with an error spectrum; it includes the Poisson noise, flux
calibration uncertainties and errors in the local pseudocontinuum level. 

After subtracting the pseudocontinuum, the fitting
 algorithm returns fluxes, EWs and FWHMs among other parameters (like the
 Gaussian $\sigma$ and peak position in $\rm \AA$) for all our
 lines. 

A similar treatment was used for the stellar absorption features CO(2-0) (at $\rm 2.29\, \mu m$) and NaI (at $\rm \lambda 2.206-2.208 \,\mu m$). Fluxes and EWs of the former were calculated by using the spectroscopic index definition of Kleinmann \& Hall (1986). The NaI line features were measured by using the index definition from F$\ddot{o}$rster Schreiber (2000). In the bottom panels of Fig.\,\ref{fig:stelmaps} we present maps of
the CO(2-0) feature; the weak NaI line is not suitable for mapping but for
integrated region analysis only. Given the asymmetric shape of the CO(2-0)
band and the doublet nature of the NaI absorption, we have not calculated FWHMs
for these features. We will leave that for Paper\,II, where the use of stellar
libraries and spectral synthesis methods will be applied to derive the
stellar kinematics.

\subsection[]{Flux, EW and FWHM Maps}\label{ss:subs14}
\begin{figure*}
\includegraphics[scale=1.0, angle=90]{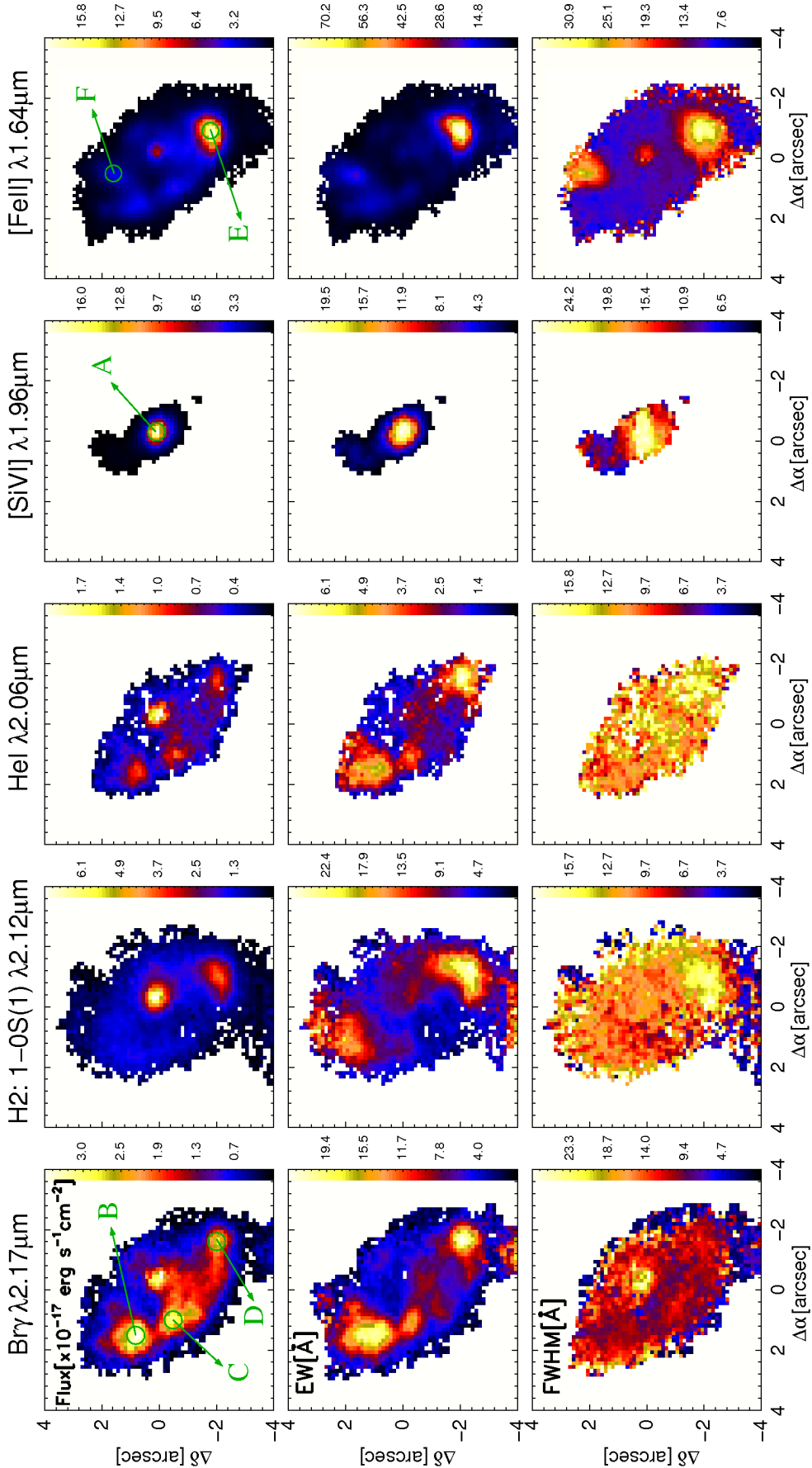}
\caption{\label{fig:maps}\small Gas emission line maps of NGC\,5135. White
  regions are those with unreliable detections for individual spaxels. From
  left to right: Br$\gamma$, H$_2$ (1-0\,S(1) transition at $\rm 2.12\, \mu
  m$), HeI$\rm \lambda 2.06\,\mu m$ [SiVI]$\rm \lambda 1.96\, \mu m$ and
  [FeII]$\rm \lambda 1.64\, \mu m$. From top to bottom: fluxes (non-corrected
  for internal extinction), EW and FWHM. The angular size
of the spatial pixels corresponds to $\rm 0.125 \times 0.125\, arcsec^{2}$. The arrows in
  the flux maps show the location of our six selected regions {\it A}-{\it F}
  and the circles represent the selected aperture sizes. At the
  distance of NGC\,5135, the $\rm 8^{\prime \prime} \times 8^{\prime \prime}$
  FoV corresponds to the central $\rm 2.3\,kpc \times 2.3\,kpc$. In all
  panels, North is up and East is left.}
\end{figure*}

In Fig.\,\ref{fig:maps} we present maps for the different phases of the interstellar medium (hereafter ISM) including the emission gas lines
Br$\gamma$ (at $\rm 2.17\, \mu m$), the H$_2$ transition 1-0\,S(1) (at $\rm
2.12\, \mu m$),  HeI$\rm \lambda 2.06\,\mu m$, [SiVI]$\rm \lambda 1.96\, \mu
m$ and [FeII]${\rm \lambda1.64\,\mu m}$, including fluxes (none are corrected for internal extinction, see Sec.\,\ref{sss:subs232}), EWs
and FWHMs (both in $\rm \AA)$. In a similar way, Fig.\,\ref{fig:stelmaps}
shows the stellar components traced by {\it H}- and {\it K}-band continua and
the CO\,(2-0) band at $\rm 2.29\,\mu m$ (both flux and EW). All maps are
centered close to the nuclear peak of the {\it K}-band continuum. At the distance of
NGC\,5135, our $\rm 8^{\prime \prime} \times 8^{\prime \prime}$ FoV
corresponds to the central $\rm 2.3\,kpc \times 2.3\,kpc$. The signal-to-noise
(hereafter S/N) per $\rm \AA$ (per
spaxel) is typically above 50 at the peaks of the emission, while in the
diffuse regions it drops to about 20. The white areas on the maps
(Fig.\,\ref{fig:maps} and bottom panels of Fig.\,\ref{fig:stelmaps}) donnot
present enough signal for reliable detections in individual spaxels.

\subsection[]{Selected Regions for study: Integrated spectra}\label{ss:subs15}
To better understand the processes taking place in the
nucleus and neighbouring regions, we have chosen six separate regions for a more
meticulous study. They have been labeled as {\it A}, {\it B},
{\it C}, {\it D}, {\it E} and {\it F}, and their locations are indicated in the first
row of Fig.\,\ref{fig:maps}. We refer the reader to Sec.\,\ref{ss:subs23} for
a detailed description of the reasons behind this selection. In
Fig.\,\ref{fig:maps}, the circles' diameters correspond to the seeing of the observations within which
we integrate our spectra to improve our S/N ratio. Such a S/N enhancement
makes it possible to run certain tests and models with weak emission features
(e.g.\ NaI, 2-1\,S(1)); this would be unfeasible on a spaxel-by-spaxel basis. At the distance of NGC\,5135, we are
integrating within aperture diameters of 157\,pc in {\it H}-band and 180\,pc in
{\it K}-band. 

To extract the integrated spectra, we have used the {\tt IDL} routine
{\tt APER}, available in the {\tt IDL} Astronomy Library. This task allowed us to obtain
fluxes from circular apertures by polynomial interpolation of the squared
pixels. Aperture uncertainties were also provided by this program by using a
close neighbours technic. As we did for the data-cube individual spaxels, we run our
Gaussian fitting program on these integrated spectra, obtaining fluxes, EWs,
FWHMs and other parameters, this time including the aperture error as an extra
source of uncertainty. In Table\,\ref{tab:lineparam} we present fluxes, EWs,
FWHMs and their $1\,\sigma$ errors for each of our spectral features
corresponding to our six selected regions. In general, the flux uncertainties were between {10--20\,\%}, although weak features (e.g.\
2-1\,S(3)) have larger uncertainties. Typically, the errors were dominated by Poisson
uncertainties. However, for lines in the blue extreme of {\it K}-band
(Br$\delta$, 1-0\,S(3), [SiVI]) the flux calibration and pseudocontinuum level
determination are more uncertain given the variability and lower transmission of
this spectral region.

Finally, we performed internal flux extinction corrections for our
measurements in regions {\it A}-{\it F}. Mean extinction values were applied
for the corresponding aperture spaxels derived from the extinction map (see Fig.\,\ref{fig:av} and Sec.\,\ref{sss:subs232} for details). We present the corrected fluxes for all our lines in Table\,\ref{tab:lineparam} with the mean $A_V$ extinction applied for
each region. It is clear from the table that some of the uncertainties from the
extinction corrections are relatively large compared to the uncorrected
flux errors. In the extreme cases, the original $<$10\% flux uncertainty of a strong line
becomes an about 30\% error in the corrected measurement. Therefore, we
decided to use fully or non-corrected fluxes according to these criteria:
Corrected fluxes were used when luminosities and other absolute flux dependant
quantities (like hydrogen column densities, $N_H$) are calculated. Also, for
the flux ratio [FeII]/Br$\gamma$ we make use of fully-corrected
fluxes for both lines. Non-corrected fluxes were only used for flux-ratio
plots between lines in the {\it K}-band.

The reason for the criteria described above is to minimise the uncertainties
whenever possible. For the flux-ratio criteria, we found that the extinction
correction between the bluest and reddest lines in {\it K}-band differs by
6-8\% depending on the {\it A}-{\it F} regions. In a flux ratio, this
uncertainty is negligible compared to the individual flux errors in both
lines. Therefore, while we do not significantly improve our measured ratio, we
increase our error bars by about a factor of 3. When [FeII]$\rm \lambda 1.64\,\mu m$
is involved, however, the situation is different. In a similar
analysis, we found that the extinction correction between [FeII] and a {\it
  K}-band line can easily differ by more than 10\%, even reaching 20\%
differences in the worst cases. These corrections are comparable to or larger
than the errors of the non-corrected flux ratios. Therefore, when [FeII] flux
is involved, internal extinction must be taken into account.

\section[]{Results and Discussion}\label{s:s2}
In this section we present the results and analysis of the spatially resolved
spectroscopy of NGC\,5135.
The many spectral features detected provide a variety of tracers of different
physical phenomena. The most frequently used emission lines in the present work
include Br$\gamma$ ($\rm 2.16\, \mu m$, recent star formation tracer), the
H$_2$ transition 1-0\,S(1) ($\rm 2.12\, \mu m$, warm molecular gas), [SiVI]
($\rm 1.96\, \mu m$, an AGN tracer) and [FeII] ($\rm 1.64\, \mu m$, a
supernova remnant tracer). Also, other features like the CO band at $\rm
2.29\, \mu m$, the NaI doublet at $\rm \lambda 2.206-2.208\,\mu m$, HeI$\rm
\lambda 2.06\,\mu m$, Br$\delta$ and the H$_2$ transitions 1-0\,S(0),
1-0\,S(2), 1-0\,S(3), 2-1\,S(1) and 2-1\,S(3) were used during the analysis.

\subsection[]{Central Gas and Stellar Structure}\label{ss:subs23}

It is evident from the Figs.\,\ref{fig:maps} and \ref{fig:stelmaps} flux maps that they are tracing different structures within the central $\rm 2.3\,kpc$ of NGC\,5135. In general, the number and spatial location of local flux peaks are totally different from one tracer to another. Only the nuclear region of this galaxy presents flux peaks in all maps. The EW maps (Fig.\,\ref{fig:maps}, second row and Fig.\,\ref{fig:stelmaps}, bottom right panel) also differ substantially from tracer to tracer, showing local peaks usually matching those in flux. The only exceptions correspond to the nuclear region, where the AGN increases the continuum level (i.\,e.\ decreasing the EW values), and to the NE region of the H$_2$ map, where a EW peak does not have a flux peak counterpart. The FWHM maps (Fig.\,\ref{fig:maps}, third row) are rather smooth compared to the previous two. Only the FWHM([FeII]) map presents clear structure, with two local peaks (north and south) separated by a third one in the nucleus. More detailed descriptions of some maps will be presented in the following pages, while some other enlightening comparisons between them will be used in different sections of this paper.

As mentioned in Sec.\,\ref{ss:subs15}, the chosen regions {\it A} to {\it F} will help us to understand the different processes affecting the gas and stars in this galaxy. However, they will also be useful to label some of the structural features observed in Fig.\,\ref{fig:maps} flux maps. Therefore, a brief description of these regions follows. Region {\it A}
corresponds to the nuclear peak in the {\it K}-band continuum. This region
includes the AGN nucleus; all emission lines measured and the continuum have a
peak at this location. Regions {\it B, C} and {\it D} correspond to bright
peaks on the Br$\gamma$ flux map, where probably very recent star formation has taken
place. Region {\it E} corresponds to the most prominent peak on the
[FeII] flux map, presumably indicating the strong effects of shocks from
supernova (SNe from now on). Also, this region matches the position of the secondary peak
observed in the 1-0\,S(1) transition of H$_2$. We remark that the peaks of
regions {\it D} and {\it E}, despite being close ($0.56^{\prime \prime}$), are
not coincident. In terms of their integrated spectra, {\it D} and {\it E} share
only a few common flux spaxels (equivalent to 1-2 spaxel's area) in the
{\it K}-band ($0.62^{\prime \prime}$ aperture), while their {\it H}-band spectra are independent because of the smaller aperture used ($0.54^{\prime \prime}$). Any PSF-wing contamination effects were found to be negligible within these two apertures. Finally, region {\it F} corresponds to a local peak on both the [FeII] flux and FWHM maps. It is also coincident with an extended
structure in [SiVI]. In Fig.\,\ref{fig:specK} and \ref{fig:specH} we show
the extracted {\it K} and {\it H}-band spectra from these six regions.

\subsubsection[]{Coronal Gas}\label{ssss:subs231a}

The [SiVI]$\lambda 1.96 \mu$m and [CaVIII]$\lambda 2.32 \mu$m coronal lines
are observed in the data, the former being our main AGN tracer
(Fig.\,\ref{fig:maps}, fourth column; e.g.\,Maiolino et al.\,2000, Rodr\'iguez-Ardila et al.\,2006a). The [CaVIII] line is too faint and too close to CO\,(3-1) to be
measured reliably. The ionization potential required to produce the [SiVI] line
(167\,eV) is usually associated to Seyfert activity where the gas is
excited just outside the broad line regions of AGNs. 

 As we can see in the Fig.\,\ref{fig:maps} flux map, the [SiVI] line traces the galaxy nucleus and it also presents a weaker ``plume'' to the NE.
This particular region, in terms of spatial scales, is at least $\rm 4\times$
larger ($\rm \approx 600\,pc$ in this case) than previous reports on
different Seyfert galaxy samples (e.g.\ Prieto et al.\ 2005, Rodr\'iguez-Ardila et al.\ 2006a) where [SiVI] structures have at the most $\rm \approx 150\,pc$
across. This finding, together with evidence for a similar structure to the SW (see Sec.\,\ref{ss:subs242}), suggest that we have detected, for the first time, the presence of ionizing cones in NGC\,5135. According to
Rodr\'iguez-Ardila et al.\ (2006a), the morphology of [SiVI] and other coronal
gas is preferably aligned with the direction of the traditional
lower-ionization cones (i.\,e.\ traced by [OIII]) seen in Seyfert galaxies.

\subsubsection[]{Ionized Gas} \label{ssss:subs231b}

The Br$\gamma$ and HeI maps (Fig.\,\ref{fig:maps}, first and third 
columns) show emission dominated by three extra-nuclear sources at
$\approx 2^{\prime \prime}$ from the galaxy center, corresponding to regions
{\it B}, {\it C} and {\it D}. As other recombination lines,
the Br$\gamma$ feature is usually associated to recent star formation, where
photoionization from massive young stars is taking place. The higher
ionization energy of HeI (24.6\,eV) compared to hydrogen (13.6\,eV) makes it a good
tracer of the most massive O-B stars and therefore, in principle, of the
youngest populations in starforming regions. An unambiguous interpretation of the HeI line, however, is rather controversial. Different authors
have claimed that this emission depends on the nebular density, temperature and
dust content, apart from the He/H relative abundance and relative ionization
fractions (Shields 1993, Doherty et al.\ 1995, Lumsden et al.\
2001). Therefore, without detailed photoionization models, certain frequently
used tests like the HeI/Br$\gamma$ ratio are very uncertain when analysing
the hottest star temperature and tracing this population in $\HII$
regions. This is why we will use HeI only as a first order spatial
identification of the youngest stellar populations. The fluxes, EWs and FWHMs 
of this feature are included in Table\,\ref{tab:lineparam} for completeness.

By comparing the Br$\gamma$ and HeI flux maps, it is evident that both are
tracing the same structures, particularly the three extranuclear peaks
mentioned above ({\it B}, {\it C} and {\it D}). It is worth noticing, however,
that the strongest peak of the HeI-emission corresponds to the central region {\it A}, as it is also the case of Br$\gamma$ after extinction correction (see
Table\,\ref{tab:lineparam}). These probably reflect that processes associated
to the AGN (e.g.\ shocks, photoionization, X-ray emission) contribute to 
 the observed fluxes. However, given the size of our aperture (180\,pc), we
cannot discard the contribution of very recent starburst to further ionize the
AGN vicinity. Gonz\'alez-Delgado et al.\ (1998) have shown for a small sample
of local Seyfert\,2 galaxies that, even in UV/optical wavelengths (highly
affected by extinction), starforming knots are clearly identified at distances
well below 100\,pc from the nucleus (see also Cid Fernandes et al.\ 2004 and
Davies et al.\ 2007).  

>From higher resolution (HST/NICMOS) Pa$\alpha$ maps (Alonso-Herrero et al.\ 2006a) we know that at least regions {\it B} and {\it C} have a much more complex structure than observed here, and are composed of a few individual knots of star formation. At least
part of the important diffuse emission component observed in the Br$\gamma$
map presents some structure in the Pa$\alpha$ counterpart, particularly within
the ``triangle'' formed by {\it A}, {\it C} and {\it D}.

\subsubsection[]{Partially Ionized Gas}\label{ssss:subs231c}

Different astronomical objects are capable of emitting [FeII]$\rm \lambda  1.64\, \mu m$, including $\HII$ regions, SNe remnants (hereafter SNRs) and AGNs, meaning that this line can be excited by different processes (see Sec.\,\ref{ss:subs24} for details). 


Strong [FeII]$\rm \lambda  1.64\, \mu m$ emission is produced by free electron
collisions in hot gas ($10^{3}$-$\rm 10^{4}\, K$). Its low ionization energy
(16.2\,eV), however, indicates that this line is strong only in partially ionized regions; otherwise the Fe would be in higher ionization states. The SNR shocks
are a suitable mechanism to produce partially ionized zones and therefore the [FeII] flux has been frequently used as a SNe rate
indicator (e.g.\ Colina 1993, Alonso-Herrero et al.\ 2003, Labrie \&
Pritchet 2006, see Sec.\,\ref{ss:subs28} for more references; see also Vanzi
et al.\ 1996 for a different view). 
It is possible, however, that X-ray photoionization
(e.g.\ from AGN or binary stars) produces a similar effect to that of shocks in the surrounding gas (Maloney et al.\ 1996).

Our [FeII] flux map is the most complex in structure (Fig.\,\ref{fig:maps}, fifth column), where two strong and three weaker extranuclear peaks
  are observed. The most prominent one corresponds to our {\it E} region and it
  seems to be a hot spot for SNe explosions given its high FWHM ($\rm \approx
  513\, km\,s^{-1}$). On the other hand, our region {\it F} is also associated
  to a high FWHM zone but without a particularly large [FeII] emission in
  comparison to the other extranuclear peaks. This region seems to match with
  the [SiVI] ``plume'' mentioned above. Developing more detailed kinematical
  arguments is beyond the scope of this paper and we will leave such analysis for Paper\,II. In Sec.\,\ref{ss:subs241} we will contrast the FeII with
  Br$\gamma$ fluxes to shed some light on the dominant mechanisms responsible
  for heating the gas in the different regions.

\subsubsection[]{Warm Molecular Gas}\label{ssss:subs231d}

The observed H$_2$ features in the spectra (Fig.\,\ref{fig:specK}) are formed by roto-vibrational transitions of the molecules. Different mechanisms
have been proposed as external energy sources including UV-fluorescence,
thermal collisional excitation by SNe fast shocks and X-rays. 

Our strongest
warm molecular gas tracer, 1-0\,S(1), shows a very different structure than
previous indicators (Fig.\,\ref{fig:maps}, second column). Two intensity peaks
dominate the emission: the strongest at the nucleus and the weaker peaking in
region {\it E}, matching the main peak in [FeII] flux. Both peaks are broad
(the later, spatially resolved), but the one in {\it E} is the dynamically
hotter as shown in the FWHM map. The remaining H$_2$ flux has a diffuse
appearance at our spatial resolution.

\subsubsection[]{Giant and Supergiant Stellar Component}\label{ssss:subs231e}

\begin{figure}
\includegraphics[scale=0.48]{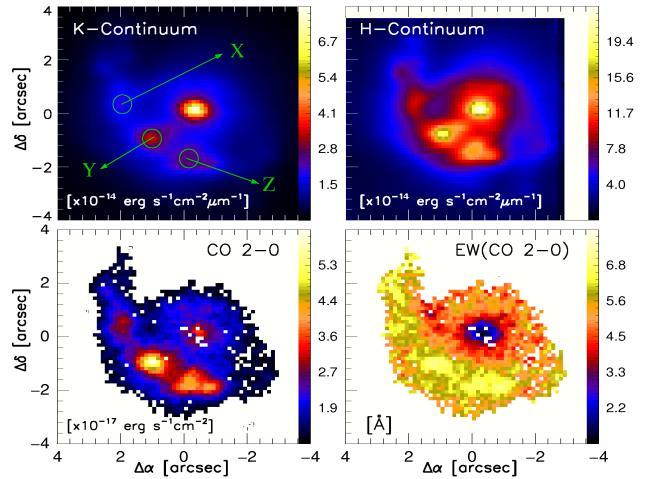}
\caption{\label{fig:stelmaps}\small Stellar emission traced by different
  indicators. White regions are those with unreliable measurements of CO in
  individual spaxels. Top panels: SINFONI {\it K}- and {\it H}-band continuum
  flux maps, the former indicating selected regions {\it X}, {\it Y} and {\it Z}. Bottom: Flux of CO\,(2-0) band at $\rm 2.29\,\mu m$ (left panel) and
  its EW (right panel). The angular size of the spatial pixels corresponds to $\rm 0.125 \times 0.125\, arcsec^{2}$.}
\end{figure}

In young starburst galaxies, massive ($\rm < 40\,M_{\odot}$) red giants and
supergiants with ages between 10-100\,Myr dominate the integrated NIR
continuum emission. As the population evolves with time, less massive giants
start dominating the emission. Few Gyr later, the IR and bolometric
luminosities are dominated by the low mass giants near the tip of the Red Giant
Branch (Renzini \& Buzzoni 1986, Chiosi et al.\ 1986, Origlia \& Oliva
2000). Spectral features such as the CO bands at $2.29\,\mu$m and the NaI doublet (at $2.206$ and $2.208\,\mu$m) are typical of K and later stellar types and also trace the luminous red giant and supergiant populations. The bottom panels of Fig.\,\ref{fig:stelmaps} show the flux and EW of the CO$\lambda2.29\,\mu$m band. White regions are those with unreliable detections of CO in individual spaxels.

In Fig.\,\ref{fig:stelmaps} top panels, the {\it H}- and {\it K}-band continuum
maps show a different spatial distribution than all previous emission line
maps in Fig.\,\ref{fig:maps}. We observed three extra-nuclear peaks which donnot correspond to any of our {\it A}-{\it F} regions. However,
from higher spatial resolution {\it J}-band continuum data (HST/NICMOS,
Alonso-Herrero et al.\ 2006a) we know that these regions have a much more
complex spatial structure with a series of knots. In any case, the strong
diffuse emission component observed in their {\it J}-band continuum is
also present in our SINFONI NIR data.

In general, from the four panels of Fig.\,\ref{fig:stelmaps} we see that all
these indicators are essentially tracing the same structures,
particularly in the extra-nuclear zone. The only
exception occurs in the nuclear {\it A} region, where the strong NIR-continuum
emission does not present a comparable counterpart in CO$\lambda2.29\,\mu$m
flux map. 

In Sec.\,\ref{ss:subs26} we will go further on characterising the stellar
population and its spatial structure by using these tracers. We will also
estimate its age in different regions by using simple stellar population
models (SSP models from now on).

\subsubsection[]{Overall Picture}\label{ssss:subs231f}

\begin{figure}
\includegraphics[scale=0.64]{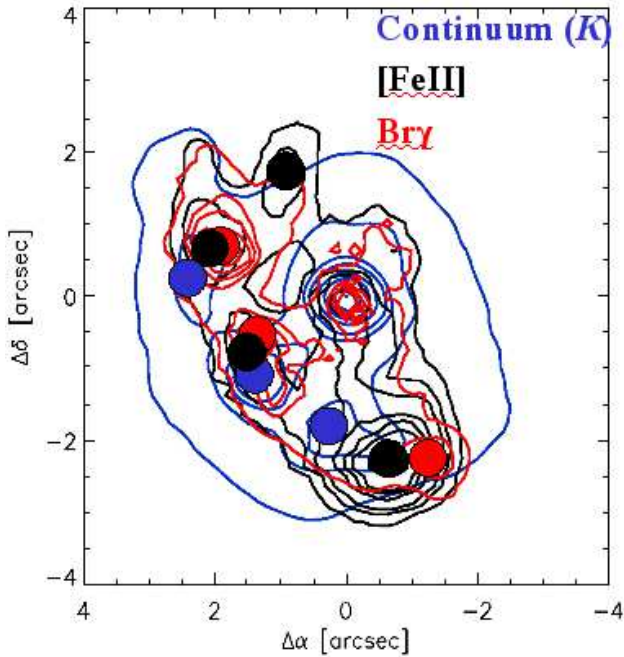}
\caption{\label{fig:contours}\small Three contour maps overplotted: {\it
    K}-band continuum (blue), [FeII] (black) and Br$\gamma$ (red). In the same
    colors, local peaks of each map are marked by circles of approximately the same size
    than our extraction apertures.}
\end{figure}

Summarising, all the structural components described in this section clearly show
different spatial distributions through their associated spectral feature
maps. We wish to highlight that, apart from region {\it A} and region {\it E}
(the [FeII]-main-peak / H$_2$-secondary peak pair), many of the other line peaks
are not spatially coincident with each other. We show this in
Fig.\,\ref{fig:contours}. Three contour maps are overplotted including
Br$\gamma$, [FeII] and {\it K}-band continuum, where the circles represent the
position of the peaks for each map, having approximately the diameter of our
extraction apertures. According to StarBurst99 models (hereafter SB99,
Leitherer et al.\ 1999), these spectral features and {\it K}-band emission
might be tracing stellar populations on different evolutionary stages,
pointing towards ages of $\approx$ 6\,Myr, $\leqslant$40\,Myr (assuming a SNR
origin of [FeII]-emission) and $\geqslant$200\,Myr, respectively. In
Secs.\,\ref{ss:subs241} and \ref{ss:subs26}  we present a more detailed
analysis of the stellar populations by using Br$\gamma$, the
CO$\lambda2.29\,\mu$m band, NaI$\lambda 2.21 \mu$m and NIR-continuum. See also
Secs.\,\ref{ss:subs241} and \ref{ss:subs28} for some discussion concerning the
origin of [FeII]-emission in this galaxy.
   
\subsection[]{Internal Extinction Map}\label{sss:subs232}

Large amounts of dust and gas are usually found in local (U)LIRGs, hiding an
important fraction of their star formation and AGN activity (Sanders et al.\
1991, Sanders \& Mirabel 1996, Solomon et al.\ 1997). Dust grains are mainly
responsible for the absorption of UV/visual photons and re-emitting them as
sub-millimeter and FIR radiation. In the NIR range, dust also absorbs
and scatters radiation, but its effects are much smaller than at shorter
wavelengths.

For NGC\,5135, the effects of internal extinction become apparent in
Gonz\'alez Delgado et
al.\ (1998) work. By comparing their optical (606W at $HST$/WFPC2, their
Fig.\,1) and UV-continuum maps ($HST$/FOC, their Figs.\,5 and 6b) we found that
our nuclear region {\it A} (centered at $+1.7^{\prime \prime}$N,
$+0.4^{\prime \prime}$W in their coordinate system) almost disappear in the UV
compared to visual, while starforming regions like {\it E} retain their
morphologies in both wavelengths. This suggests that the AGN neighborhood is
more obscured than other circumnuclear regions of this galaxy.

\begin{figure}
\includegraphics[scale=0.47]{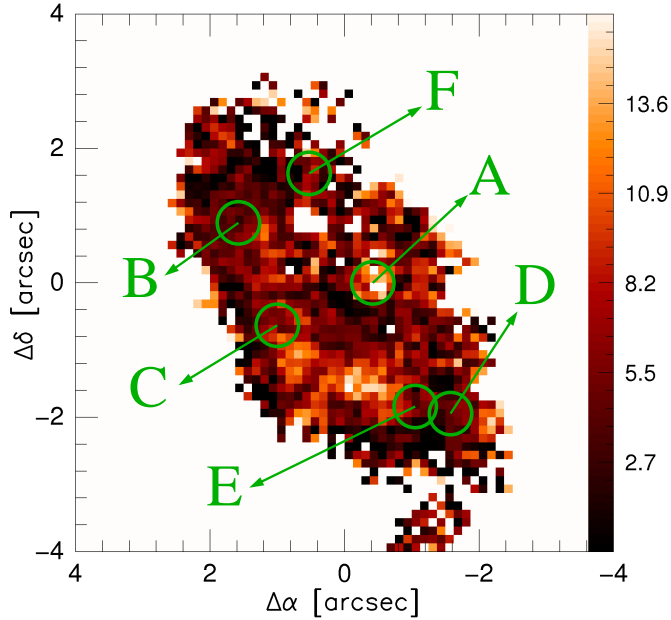}
\caption{\label{fig:av}\small $A_V$ gas extinction map of NGC\,5135 based on
  Br$\gamma$/Br$\delta$ line ratio. The arrows show our six selected regions
  {\it A}-{\it F} and the circles approximately represent the apertures in {\it
  K}-band. White regions are those with unreliable $A_V$ measurements in
  individual spaxels. The Calzetti et al.\ (2000) extinction law has been assumed.}
\end{figure}

We can estimate the internal gas extinction in NGC\,5135 by using
observations of different hydrogen recombination lines. We adopt our own 
Br$\gamma$ and Br$\delta$ measurements for this purpose. The theoretical ratio
between these two lines (Br$\gamma$/Br$\delta$ = 1.52 at $T=10\,000\,\rm K$ and
$n_e = 10^4 \,\rm cm^{-3}$, Osterbrock 1989) was compared with the measured
fluxes for each spaxel. Therefore the extinction in magnitudes ($A_{\lambda}$)
for both lines can be combined in the form

$$
A_{Br\gamma} - A_{Br\delta} =
-2.5\cdot\Big[\log\Big(\frac{F_{Br\gamma,O}}{F_{Br\gamma,T}}\Big) -
\log\Big(\frac{F_{Br\delta,O}}{F_{Br\delta,T}}\Big)\Big] 
$$
\begin{equation}\label{eq:av_calc}
= -2.5 \cdot \log\Big[\frac{(F_{Br\gamma}/F_{Br\delta})_O}{(F_{Br\gamma}/F_{Br\delta})_T}\Big]
\end{equation}

where $F_{\lambda,O}$ and $F_{\lambda,T}$ are the observed and theoretical
fluxes for a line centered at $\lambda$, respectively. Then, by interpolating
the extinction law of Calzetti et al.\ (2000), we put eq.\,\ref{eq:av_calc} in
function of $A_V$ for each spaxel ($A_{Br\gamma}=0.096\,A_V$ and
$A_{Br\delta}=0.132\,A_V$). From this expression we obtain the optical gas
extinction map presented in Fig.\,\ref{fig:av}. White regions represent those where none flux information is available or where unreliable ratios were estimated. We remark that these extinction values are lower-limits only. This is because the gas might be optically thick (contrary to what is implicit in the theoretical values used above), therefore hiding the deeper regions from us. In terms of {\it individual spaxels}, the $1\,\sigma$ uncertainties in this
$A_V$-gas map varies from 10-30$\,\%$ to up to 80$\,\%$. The former typically
corresponds to high S/N zones, like regions {\it A}-{\it F}, while
the later matches with peripheral areas and some inter-region zones where the
S/N is low. 

As we can see from Fig.\,\ref{fig:av}, the regions with higher $A_V$ values
are:

{\it (i)} The Nuclear Region: The average visual
extinction is about 10\,mag (0.9\,mag at 2.2\,$\mu$m), with a local peak
reaching the maximum of our scale at about 17\,mag. This result is consistent
with the conclusions of Gonz\'alez Delgado et al.\ (1998). At least
qualitatively, it is also in agreement with findings of Levenson et al.\
(2004) using {\it Chandra} X-ray data. They claim that the nuclear AGN engine
must be obscured by a high hydrogen column density ($N_H \rm \geq 10^{24}\,
cm^{-2}$, having in mind they have used $4\times$ larger apertures than
ours). Although we do not find such high $N_H$ values (see
Sec.\,\ref{sss:subs252}, Table\,\ref{tab:colden}), our $1$-$\rm 3\times
10^{22}\,cm^{-2}$ measurements in region {\it A} are by far the highest of our
{\it A}-{\it F} apertures. 


{\it (ii)} A region south from the nucleus ($\Delta\alpha \approx
[-0.5,1.0]$, $\Delta\delta \approx [-2.0,-1.0]$, approx.): It has a mean visual extinction of 10\,mag. Presenting some local peaks of about 17\,mag, this area covers the space between our regions {\it C} and {\it E}. No strong emission in
Br$\gamma$, Br$\delta$ and HeI flux maps is observed in this area, but only a
weaker, smooth component at SINFONI's resolution. Despite the fact that both Brackett lines are clearly detected in this zone, we cannot discard that some of the peaks might be partially produced by flux uncertainties, particularly of the
Br$\delta$ line and its complex local continuum determination. A higher
continuum level estimation would decrease this line flux,
therefore, artificially increasing the extinction (see
eq.\,\ref{eq:av_calc}). In any case, because this region does not match with
our {\it A}-{\it F} zones, it will not affect our extinction corrected fluxes
(Table\,\ref{tab:lineparam}) and conclusions derived from them. 

The rest of the map presents a median visual extinction of 6\,mag which
translates into 0.55\,mag at 2.2\,$\mu$m.

Alonso-Herrero et al.\ (2006a) presented an F110W-F160W color map of
NGC\,5135 (using NICMOS-NIC2; similar to {\it J}-{\it H} color) which can be
interpreted as a stellar extinction map to a first order. It is difficult,
however,  to directly translate this map to $A_V$ extinction to the stars, mainly because we have to assume an age for the stellar
population (to get its theoretical {\it J}-{\it H} color) in a galaxy which
may have a composite stellar population (see Secs.\,\ref{ss:subs241} and
\ref{ss:subs26}). In terms of their structure, a comparison between
Fig.\,\ref{fig:av} and Alonso-Herrero et al.\ color map shows that they are
remarkably different. While both maps seem to agree in having the larger
extinctions around the nucleus, the color map clearly traces the
nuclear spiral arms while no sign of this is observed in our $A_V$ map. We do
not think that the higher spatial resolution of NICMOS data can totally
explain such a radical mismatch by itself. Instead, the close wavelength
proximity between our two Brackett emission lines might be the responsible of
the apparent $A_V$ map lack of structure. The similar extinctions for both
Brackett emission lines produce $A_V$ values (from eq.\,\ref{eq:av_calc}) very
sensitive to small flux uncertainties. Therefore, the detailed structure in
the $A_V$-gas map gets easily lost within Br$\gamma$ and Br$\delta$ errors. We
remark, however, that for larger areas (sampled by few tenths of spaxels
like regions {\it A}-{\it F}), the mean $A_V$ values should be a reasonably
good estimation of the local internal extinction.

\subsection[]{Ionization Mechanisms of Gas}\label{ss:subs24}
The presence of ionized gas is usually interpreted as a tracer of recent star
formation, where UV radiation from massive O-B stars is capable of keeping
the surrounding gas in an ionized state. In the NIR, recombination lines
like Pa$\alpha$ and Br$\gamma$ are frequently used as tracers of this
activity.

 However, there are other mechanisms capable of ionizing the
ISM. Shocks produced in SNR can partially ionize the
gas. In the NIR, SNRs can be traced by the [FeII]${\rm \lambda1.26\,\mu m}$
and [FeII]${\rm \lambda1.64\,\mu m}$
lines (e.g.\ Greenhouse et al.\ 1991, Alonso-Herrero et al.\ 2003). A strong
[FeII] flux is not expected in normal $\HII$ regions (Mouri, Kawara \&
Taniguchi 2000): their abundant fully-ionized hydrogen would produce Fe atoms
in higher ionization states. Therefore, by using our Br$\gamma$ and
[FeII]${\rm \lambda1.64\,\mu m}$ fluxes we are tracing different ionization
mechanisms with also different ionization efficiencies.

In Fig.\,\ref{fig:feIIbrG_EWbrG} we present the extinction corrected
[FeII]/Br$\gamma$ flux map. This diagnostic allows to see which ionization
mechanism is more dominant in different regions of the galaxy. An
[FeII]/Br$\gamma$ ratio $\lesssim 1$ is usually taken as characteristic of a
starburst, while much larger values are associated with only partially ionized
regions by shocks (e.g.\ Graham et al.\ 1987, Kawara et al.\ 1988, Mouri et
al.\ 1990, Alonso-Herrero et al.\ 1997, 2001).

\subsubsection[]{Circumnuclear Starforming Regions}\label{ss:subs241}

\begin{figure}
\includegraphics[scale=0.47]{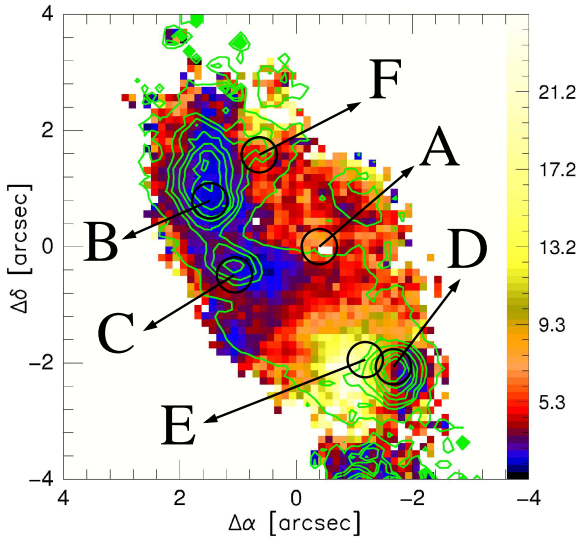}
\caption{\label{fig:feIIbrG_EWbrG}\small {Extinction corrected
    [FeII]/Br$\gamma$ line ratio map. In contours we plot the EW(Br$\gamma$) as an age indicator. The arrows show our six selected regions
  {\it A}-{\it F} and the circles approximately represent the apertures in {\it
  K}-band.}}
\end{figure}

 Fig.\,\ref{fig:feIIbrG_EWbrG} shows that only three regions (associated to
 {\it B}, {\it C} and {\it D}) seem to have a non-negligible contribution from
 starburst activity. In the same figure, we overplot in contours the
 EW(Br$\gamma$) as an upper-limit age indicator. It is clear that the
 EW(Br$\gamma$) peaks match very well those areas with low-[FeII]/Br$\gamma$
 ratio, pointing out that a young stellar population contributes to ionize the
 ISM in these areas. We have tested this idea by using the SB99 models with instantaneous star formation, solar metallicity and Kroupa initial mass function (hereafter IMF). In
 Table\,\ref{tab:age} we present the predicted ages (upper limits) of the
 stellar populations for our selected regions, derived from our EW(Br$\gamma$)
 measurements of Table\,\ref{tab:lineparam}. As we can see, {\it B}, {\it C}
 and {\it D} are the three youngest areas with $\rm \approx 6\,Myr$ old,
 while {\it E} and {\it F} are slightly older. These results are consistent
 with previous findings with UV/visual data (Gonz\'alez Delgado et al.\
 1998). It is possible that some  contamination from the nearby starburst
 regions has been introduced within {\it E} and {\it F} zones, particularly
 for the former. \footnote{The age presented for
 region {\it A} is an upper limit for two reasons: as for the other regions,
 we are assuming that all Br$\gamma$-emission comes from star formation; but
 also in region {\it A} the presence of the AGN increases the continuum level
 producing lower EW(Br$\gamma$) values and therefore, larger ages than
 expected for a pure starburst area.} 
   
\begin{table}  
\begin{center}
 \caption{StarBurst 99 Ages from EW(Br$\gamma$)\label{tab:age}} 
 \begin{tabular}{@{}|l|c|@{}}
\hline
Reg     &  Age$^a$ [Myr]  \\
\hline
{\it A} & $\leq 7.02$ \\
{\it B} & $6.15_{-0.03}^{+0.02}$ \\
{\it C} & $6.38_{-0.43}^{+0.03}$ \\
{\it D} & $6.14_{-0.03}^{+0.02}$ \\
{\it E} & $6.52_{-0.03}^{+0.04}$ \\
{\it F} & $6.51_{-0.05}^{+0.04}$ \\
\hline
\end{tabular}\\
\end{center}
\footnotesize{($^a$) Ages with their $\pm 1\sigma$ errors. Ages should be
  considered as upper limits only. See text for details and footnote 3 for
  region {\it A}.}
\end{table}

A plausible explanation for the local [FeII]/Br$\gamma$ variations might be suggested from
Table\,\ref{tab:age} ages, where regions {\it E} and {\it F} are marginally
older than the others. According to models, after about $\rm 6$-$\rm 7\,Myr$ important changes start taking place in young stellar
populations. The massive O-B stars responsible for most of the Br$\gamma$-emission explode as SNe, diminishing this flux feature in the integrated
spectra. On the other hand, after $\rm \sim 4\,Myr$ since the starburst, the
SNe rate has been increasing dramatically, enriching the ISM with
Fe. Therefore, two different processes radically change the [Fe/H] abundance and
the Br$\gamma$ flux: when total ionization from O-B stars leaves place to partial ionization from SNR shocks, the Fe-enriched ISM increases the
[FeII] flux, while the decline of O-B stars decreases Br$\gamma$. If a certain region
has not reached yet this crucial stage on its stellar evolution (like {\it B}, {\it C} and {\it D}) we would observe low
[FeII]/Br$\gamma$ ratios. On the other hand, if the stellar population has
reached this point (like in {\it E} and {\it F}), much higher [FeII]/Br$\gamma$
ratios would be observed.

The high values of [FeII]/Br$\gamma$ ($>$ 20) in regions {\it E} and
{\it F} are, in fact, very similar to those observed in SNRs, e.g.\ $\sim
30$ in IC\,443 (Graham et al.\ 1987) or 34 in RCW\,103 (Oliva et al.\ 1989).
In the FWHM([FeII]) map of Fig.\,\ref{fig:maps} we see that both sources are the
only ones with very high values (or large velocity dispersions). This would be, in principle, in excellent agreement with a scenario where SNR shocks have
perturbed the kinematics of these areas.

\begin{figure}
\includegraphics[scale=0.6]{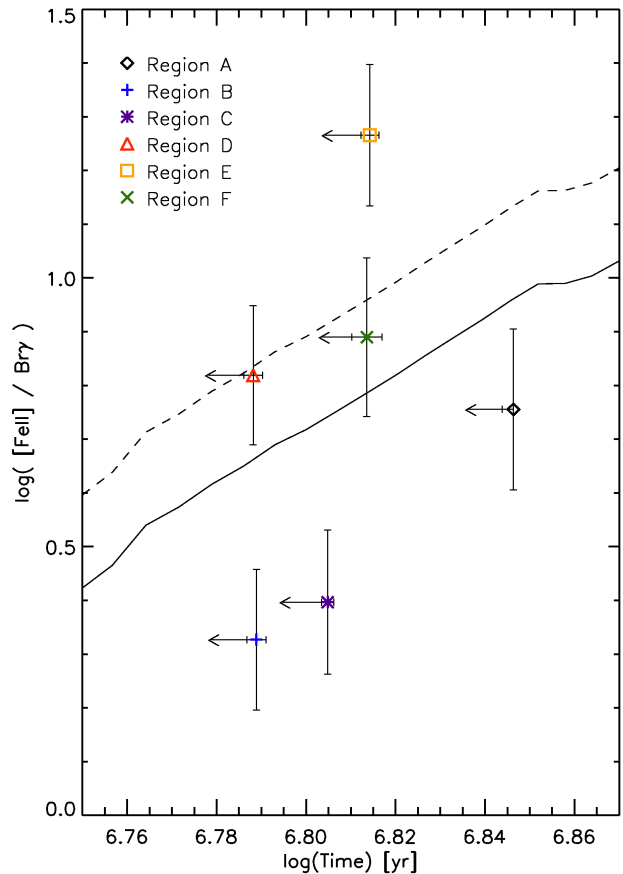}
\caption{\label{fig:SB99_fe-brG}\small {[FeII]/Br$\gamma$ vs.\ Age (upper
    limits from EW(Br$\gamma$)) for our six
    regions. Based on SB99 models, different relations have been used to derive the [FeII] flux from the SNe rate: dashed line, Calzetti (1997); solid line, Alonso-Herrero et al.\ (2003).}}
\end{figure}

We have compared all our [FeII]/Br$\gamma$ measurements with predictions from the SB99 models. On its latest versions, SB99 does not give directly the [FeII] fluxes but the SNe rate. So we transform these values to [FeII] flux by using empirical relations from Calzetti (1997) and Alonso-Herrero et al.\ (2003). By using these expressions, we can further test (to first order) the SNR scenario. See Sec.\,\ref{ss:subs28} for further discussion about SNe rate estimations. In Fig.\,\ref{fig:SB99_fe-brG}
we present SB99 predictions with the two empirical relations mentioned
above. We overplot our measured [FeII]/Br$\gamma$ ratios for {\it A}-{\it F}
with their uncertainties. We have used the upper-limit ages of
Table\,\ref{tab:age} to fix the abscissa of our data points. 

Only regions {\it D} and {\it F} show a clear agreement with the SNR models at those ages. Region {\it D} has an important [FeII] emission contribution from the
broad peak traced by region {\it E}. So, the agreement between this young
starburst region and the models is not surprising. The good match with {\it F}
only reinforces the link between this area and SNRs as mentioned before using
kinematic arguments. The starburst {\it B} and {\it C} regions lay below model predictions,
pointing towards pure stellar photodissociation mechanisms as the responsible
of the gas ionization. 

Certainly, the most interesting outlier is region {\it E}, where the models predict a smaller
[FeII]/Br$\gamma$ ratio than observed. This implies that an additional
ionization source must be invoked to explain the observations at the center of
this resolved [FeII] peak (region {\it D}, taken as a peripheral sample of this
peak, does not need such an extra ionization source). A plausible
alternative comes from  X-ray photoionization. As mentioned in previous
sections, Levenson et al.\ (2004) have published {\it Chandra} X-ray data for
NGC\,5135. They have identified in their total band map ($0.4$-$\rm 8\,keV$) a
strong emission corresponding to region {\it E}. According to these authors,
the non-thermal component (power law and the bulk of the hard X-ray emission)
is likely to be associated to X-ray binaries in the starburst, while the
thermal one (most of the soft emission) is typical of a starburst
galaxy. These mechanisms may explain the difference between models and {\it E}
data in Fig.\,\ref{fig:SB99_fe-brG}, where both, X-rays and SNR shocks are
contributing to the observed [FeII]/Br$\gamma$ ratio.

\subsubsection[]{The Ionization Cones and Central AGN}\label{ss:subs242}

The central region {\it A} of NGC\,5135 deserves a separate section given
its particular characteristics with respect to the other starforming regions. 

As mentioned in previous sections, the AGN activity has been traced by the
[SiVI] coronal line. The [SiVI]-luminosity for this region is $\rm
L_{[SiVI]}=3.6\times 10^{39}\,erg\,s^{-1}$, in agreement with similar
measurements in other local Seyfert\,2 galaxies, like Circinus, NGC\,1068,
NGC\,3227 and Mrk\,609 which span a range $\sim 10^{39}$-$\rm
10^{40}\,erg\,s^{-1}$ (the two later objects are Seyfert 1.5 and 1.5-1.8,
respectively; Rodr\'iguez-Ardila et al.\,2006a, Zuther et al.\,2007).
 
By using Br$\gamma$-emission, we can make a row estimation of the AGN contribution to the gas photoionization. If we assumed that the (extinction corrected) Br$\gamma$ flux in region {\it A} is entirely produced by the AGN, we can compare this value with the total Br$\gamma$ flux in our FoV. According to this, the AGN contributes in $\rm \lesssim 3\%$ of the gas photoionization within the central $\rm 2.3\,kpc$ of this galaxy. By doing a similar calculation, we also estimate that the AGN represents $\rm \lesssim 3$ and $\rm \lesssim 5\%$ of the NIR-luminosity in the ${\it H}$- and ${\it K}$-bands, respectively.

Concerning the ionization mechanisms, on the one hand the AGN embedded in this
zone is likely to play an important role {\it at local level} by photoionizing the gas in the nuclear neighborhood (Marconi et al.\ 1994, Ferguson et al.\ 1997, Contini et al.\ 1998,
see also Rodr\'iguez-Ardila et al.\ 2006b and references therein). The
[FeII]/Br$\gamma$ = 5.7 measured in this region is consistent with
typical Seyfert galaxy values from a compilation of different object types
shown in Alonso-Herrero et al.\,(1997).

The ``plume'' shape structure in [SiVI]-emission (NE of the nucleus,
Sec.\,\ref{ssss:subs231a}, Fig.\,\ref{fig:maps}) probably traces the direction
of a cone, pointing that the AGN is also capable to ionize the ISM at larger distances. The presence of
shocks from outflowing winds would produce such ionized plasma and
cone-features (Viegas-Aldrovandi \& Contini 1989). The [SiVI]-emission is
particularly broad in {\it A} with a FWHM of $\rm \approx 28\,\AA$ (or $\rm
\approx 430\, km\, s^{-1}$), strengthening the idea of an important
interaction between AGN outflows and the surrounding material. It is
interesting to notice that the FWHM([SiVI]) nuclear peak structure extends
approximately along the semi-minor axis as expected in outflow models of
Heckman et al.\,(1990). This may suggest the presence of multiple components
in the [SiVI] kinematics. The high FWHM(Br$\gamma$) in this area ($\rm \approx
22\,\AA$ or $\rm \approx 300\, km\, s^{-1}$) is also consistent with this
hypothesis. In Paper~II we will find out if this scenario receives further
support from our detailed kinematical study. The NE [SiVI] structure has a SW
counterpart not detected on a spaxel-by-spaxel basis. By integrating few
spaxels at $2^{\prime \prime}$ SW from the nucleus, we have detected clear
[SiVI]-emission which is absent in other areas close to region {\it A}. This
SW component is probably weaker because of extinction and projection effects.

On the other hand, in Levenson et al.\ (2004) work on X-rays, the authors have found the
  strongest emission peak in our region {\it A}, which may also contribute to
  the ionization in this area. They suggest, however, that the bulk of the
  thermal (collisional) X-ray emission, even in the nucleus, {\it is
  stellar}, not due to the AGN. In principle, this would be
  understood in terms of the larger apertures they use in {\it A} (about
  $4\,\times$ ours) within which a stronger stellar component, that we miss,
  might be introduced. However it is also true that compact starforming
  regions have been usually found in AGN's neighborhood at distances of tens
  of pc (e.g.\,Gonz\'alez Delgado et al.\,1998, Weaver 2001, Davies et al.\
  2007). Considering the significant dilution from the AGN, the later point is
  reinforced for NGC\,5135 by our findings of NaI and CO stellar absorption in
  this zone.

Therefore in region {\it A} and in the [SiVI]-cone the AGN photoionization and
outflows seem to play an
important role on ionizing the gas. However, a stellar contribution cannot be totally
discarded, particularly in region {\it A}. Then, it is comprehensible why in
Fig.\,\ref{fig:SB99_fe-brG}, the starburst-based models cannot reproduce the
observed [FeII]/Br$\gamma$ ratio of this region, in addition to the age
uncertainty in this zone. 
\\
\\
Finally, the remaining extended emission in Fig.\,\ref{fig:feIIbrG_EWbrG} has rather
large [FeII]/Br$\gamma$ ratios, typically between 3 and 7. This suggests that
the ISM gas is mainly ionized by other mechanisms (e.g.\,SNR shocks, X-rays)
instead of photodissociation from star formation.

\subsection[]{Dating the Stellar Population: Further constrains from Absorption Lines}\label{ss:subs26}

In the previous section, we have studied a young stellar population ($\rm
\leq 7\,Myr$) with either an important fraction of O-B stars or in a transition
period with high SNe rates. We attempted to understand its importance
on ionizing the ISM gas by both, the use of the [FeII]/Br$\gamma$ map
structure, and the estimation of the stellar population
age. Because of this, we now consider appropriate to introduce at this point the
analysis of the stellar population traced by the NIR-continuum and
CO$\lambda2.29\mu$m absorption maps of Fig.\,\ref{fig:stelmaps}. By doing
this, we will address the open issues presented in Sec.\,\ref{ssss:subs231e}.  

In Fig.\,\ref{fig:stelmaps}, we have seen that NIR-continuum and CO$\rm \lambda
2.29\,\mu m$ maps have a very similar structure everywhere but in region {\it A}. The
absence of a strong central peak in the CO flux map can be explained by the influence
of the AGN in the stellar continuum. The strong X-ray/UV radiation from 
AGN is reprocessed by the surrounding dust grains and re-emitted in the
IR (e.g.\ Clavel et al.\,1992, Forbes et al.\,1992). Therefore, the continuum
flux in {\it H} and {\it K}-bands increases, diluting the CO and other stellar
absorption features. The high extinction we estimate for this area (see also
Alonso-Herrero et al.\,2003) and the finding of hot dust
emission ($\sim 200\,$K) in the nucleus of NGC\,5135 (Gemini/T-ReCS {\it
  N}-band data, Alonso-Hererro et al.\ 2006b) are consistent with this
idea. The high temperatures required for the dust to emit in {\it
  K}-band, however (1200-1500\,K; e.g.\ Clavel et al.\,1989, Glass 1992,
Rodr\'iguez-Ardila et al.\,2005a, 2006b), limit this effect to rather short distances from the AGN ($\le 100\,$pc; Marco \& Alloin 1998, 2000). In fact, the observed dilution towards the center of the EW(CO) map (Fig.\,\ref{fig:stelmaps}, bottom-right panel) occurs at similar distances of about $0.4^{\prime \prime}$
($\sim \rm 120\, pc$). This apparently close agreement, however, is almost
certainly driven by PSF effects. It is possible than hot dust emission is confined to even shorter distances from the
active nucleus. In regions farther away from the nucleus
($\gtrsim 200\,$pc) the EW(CO) values are free of the AGN contamination and can be
confidently used for further tests of the stellar population properties.

As for {\it A}-{\it F} zones, we extracted integrated spectra for other three
extra-nuclear regions labeled as {\it X}, {\it Y} and {\it Z} (see
Fig.\,\ref{fig:stelmaps}, top-left panel). They do not match with our {\it
  A}-{\it F} apertures but coincide with the three extra-nuclear peaks of
the NIR-continuum (this is clearly shown in Fig.\ref{fig:contours}) where the
CO signal is higher.
Our aim is to do a
general comparison between the young and (potentially) old stellar
populations for the entire galaxy, based on individual regions where
one or the other is dominant. In this way, we base our conclusions on
higher S/N data.

We estimate the stellar population ages in {\it X}-{\it Z} by using
the EW(CO$\lambda 2.29\,\mu$m) and the models SB99 and {\tt STARS} (Sternberg
1998, Thornley et al.\ 2000, Davies et al.\ 2003). Both models include
treatments for the thermally pulsing AGB stars. However, without knowing
parameters like the metallicity and characteristics of the star formation
history, the model predictions might vary from case to case. We have run these
models using different IMFs (Kroupa, Salpeter) and metallicities (0.04, 0.02, 0.008) without finding
substantial variations which may change our conclusions. Therefore we have
assumed solar metallicity, Kroupa IMF and the Genova isochrons for all our age
predictions. Only the use of different star formation histories (instantaneous
and continuous) has a relevant impact on this respect. We will discuss this
particular point in the following paragraphs.

Strictly speaking,  {\tt STARS} makes use of exponentially decaying star formation rates (hereafter
SFRs). However, by using a long characteristic time-scale ($\rm 1000\,Gyr$) we
recover a continuous SFR, while by using a short one we
basically recover instantaneous burst results for ages $\rm \ge 1\,Myr$. We can
directly compare this models to instantaneous and continuous SFR models from
SB99 (R.\,L.\,Davies, private communication).  

\begin{figure}
\includegraphics[scale=0.55]{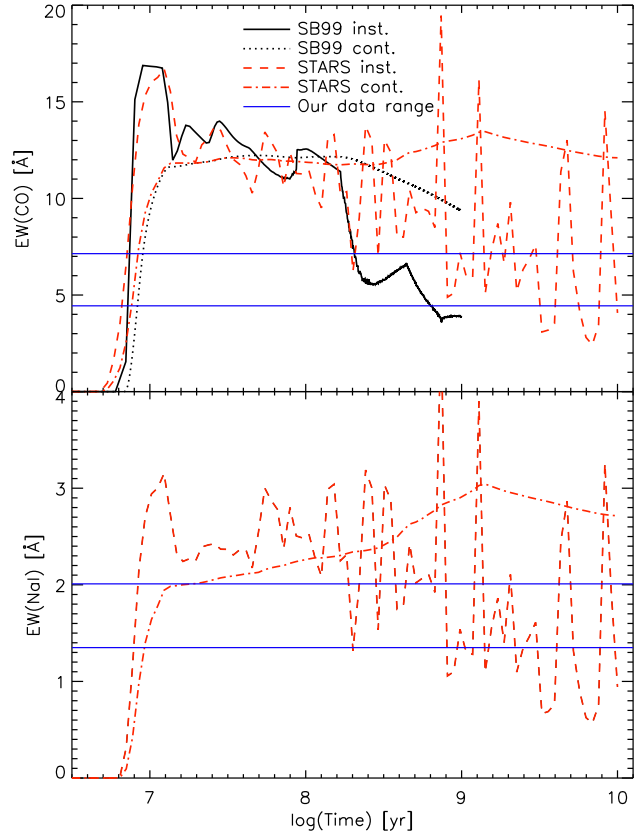}
\caption{\label{fig:SSP_models}\small {SB99 and {\tt STARS} model
    predictions for EW(CO) and EW(NaI). The two horizontal lines represent the
    range of our data for regions {\it A}-{\it F} and {\it X}-{\it Z}.}}
\end{figure}

In Table\,\ref{tab:XYZage} we present
instantaneous-burst model results for regions {\it X}-{\it Z}. Because SB99 and {\tt STARS} models interpolate the
EWs from spectral grids, their predictions become quite noisy at larger ages
($\ge \rm 100\,Myr$), particularly for {\tt STARS} were a template star
library is used (see Fig.\ref{fig:SSP_models}; R.\,L.\,Davies, private
communication). Therefore, in cases of old stellar populations ($\ge \rm
100\,Myr$) a range of ages is presented for SB99, while lower limits are shown
when {\tt STARS} is used. 

\begin{deluxetable*}{|l|ccccc||ccc||cc|}
\tablecaption{CO$\lambda 2.3\,\mu$m, NaI$\lambda 2.21 \mu$m and $\rm Br\gamma$
  EWs and Instantaneous burst model ages for regions X-Z\label{tab:XYZage}} 
\startdata
\hline
\scriptsize{Reg} & \scriptsize{EW(CO)} & \scriptsize{$\rm Age1_{SB99}^{CO}$} &
  \scriptsize{$\rm Age1_{\tt STARS}^{CO}$} & 
  \scriptsize{$\rm Age2_{SB99}^{CO}$} & \scriptsize{$\rm Age2_{\tt STARS}^{CO}$} & \scriptsize{EW(NaI)} &
  \scriptsize{$\rm Age1_{\tt STARS}^{NaI}$} & \scriptsize{$\rm Age2_{\tt STARS}^{NaI}$} & \scriptsize{EW($\rm Br\gamma$)} & \scriptsize{$\rm Age_{SB99}^{\rm Br\gamma}$} \\[0.07cm] 
          &  [\AA]         &   [Myr]                                         &      [Myr]                                      &   [Myr]      &        [Myr]    &         [\AA]       &        [Myr]                                   &        [Myr]  &       [\AA]         &       [Myr]       \\[0.07cm]                             
\hline
{\it X}   & 6.44 $\pm$ 0.4 &   $7.37_{-0.03}^{+0.03}$  &   $7.01_{-0.08}^{+0.05}$  &    213-453   &    $\ge$ 200    &    1.76 $\pm$ 0.08  &  $8.20_{-0.06}^{+0.06}$  &    $\ge$ 200  & 6.01    $\pm$ 0.64  &   $6.80_{-0.05}^{+0.06}$  \\[0.07cm]      
{\it Y}   & 7.14 $\pm$ 0.3 &   $7.42_{-0.02}^{+0.02}$  &   $7.14_{-0.05}^{+0.05}$  &    205-211   &    $\ge$ 200    &    1.99 $\pm$ 0.07  &  $8.38_{-0.06}^{+0.04}$  &    $\ge$ 200  & 7.15    $\pm$ 0.53  &   $6.70_{-0.03}^{+0.05}$  \\[0.07cm]      
{\it Z}   & 6.90 $\pm$ 0.3 &   $7.40_{-0.02}^{+0.03}$  &
  $7.10_{-0.10}^{+0.05}$  &    205-445   &    $\ge$ 200    &    1.88 $\pm$
  0.08  &  $8.29_{-0.06}^{+0.06}$  &    $\ge$ 200  & 6.64    $\pm$ 0.57  &
  $6.74_{-0.04}^{ +0.05}$  \\[0.07cm]      
\enddata
\\[0.24cm]
\end{deluxetable*}

As we can see in Fig.\ref{fig:SSP_models} (upper panel), 
instantaneous-burst models produce two possible stellar population ages
consistent with our data (around $\rm 7$ and $\rm \ge 200\,Myr$,
respectively), while continuous-star-forming ones produce a single range of
ages around $\rm 7\,Myr$. Although a stellar population of $\rm \sim 7\,Myr$
would agree with our findings in {\it B}-{\it F} zones using $\rm
EW(Br\gamma)$, the presence of a second, totally different population receives support
from two different sources.

First of all, our CO-flux (Fig.\,\ref{fig:stelmaps} bottom left panel) and
$\rm Br\gamma$-emission maps (Fig.\,\ref{fig:maps} top of first column) clearly
trace different structures (peaks) in this galaxy, usually separated by more
than $\rm 200\,pc$ in projection. On the other hand, the CO-flux map almost
mimics the structure observed in the NIR-continuum, which in turn is frequently
used as a tracer of more evolved
stellar populations than those with high $\rm EW(Br\gamma)$. Further
support for this argument comes from Alonso-Herrero et al.\,(2006a)
NICMOS/$HST$ data of this galaxy (see Alonso-Herrero et al.\,2006b, Fig.\ 1 second row for
sharper figures). The higher spatial resolution of their NIR-continuum and
Pa$\alpha$-emission maps show the peaks observed with SINFONI actually have a
complex structure. Moreover, the NICMOS continuum and Pa$\alpha$ peaks are not coincident spatially, demonstrating that, between the neighbouring
aperture pairs {\it X}-{\it B}, {\it Y}-{\it C} and {\it Z}-{\it D}, we are
picking-up different spatially resolved regions (even though some
PSF-contamination effects between aperture pairs). In terms of structure, further support to spatial variations in the stellar population comes from the EW(CO) map (Fig.\,\ref{fig:stelmaps} bottom right panel). This figure shows a clear radial (projected) distinction in function of EW. While at distances of $\rm \sim 600\,pc$ from the nucleus we find $\rm EW(CO)\sim 7\,\AA$, this value decreases to $\rm \sim 5\,\AA$ at $\rm \sim 300\,pc$. Such spatial segregation supports the idea of a change in the stellar population properties (e.g.\ a variation in the relative contribution of red giants and supergiants to the integrated light).

Secondly, the EW of different spectral features also point towards different
stellar populations in {\it X}-{\it Z} with respect to {\it B}-{\it D}
zones. On the one hand, the EW(CO) of the former are larger than those for
the later as we can see in Tables\,\ref{tab:XYZage} and \ref{tab:lineparam},
respectively. The quoted uncertainties show that these differences are
significant 
at a $4$-$5\,\sigma$ level. We observe the EW differences are
mainly driven by CO-absorption variations rather than continuum-level
changes. This suggests the existence of a different stellar population in {\it
  X}-{\it Z} with respect to {\it B}-{\it D}, the former regions having a more
important red giant/supergiant contribution to the integrated spectra
(e.\,g. Origlia \& Oliva 2000). On the other hand, as we can see in
Tables\,\ref{tab:XYZage} and \ref{tab:lineparam}, the EW(Br$\gamma$) is
substantially lower in regions {\it X}-{\it Z} than in {\it B}-{\it D} zones,
also pointing towards two different populations. Putting together all the observational evidence, the NIR-continuum and CO-absorption (red giant/supergiant stars) and
$\rm Br\gamma$-emission (O-B stars) suggest the existence of (at least) two
stellar populations in this galaxy.

 However, as mentioned above, the stellar population model predictions are not
 that clear about this point, depending on the star formation history 
 used (see Fig.\ref{fig:SSP_models}). If we assume continuous SFR models, the
 EW(CO) for all our apertures are consistent with a $\rm \sim 7\,Myr$ old
 population. By using the $\rm EW(Br\gamma)$ with the same models, however, we
 get different results. The {\it B}-{\it D} regions are now consistent with a
 $\rm 10\,Myr$ old stellar population, while regions {\it X}-{\it Z}  are
 better constrained by a $\rm \sim  60\,Myr$ old one. Clearly, the use of
 continuous models with the measured EWs produce inconsistent results between
 the two tracers. 

Instead, if we use instantaneous SFR models, the observed EW(CO)
 are consistent with two populations of $7$-$8$ and $\rm \ge 200\,Myr$ old,
 respectively (see Table\,\ref{tab:XYZage} and \ref{tab:B_Fage} for regions
 {\it X}-{\it Z} and {\it B}-{\it F}, respectively). This degeneracy is broken by using $\rm EW(Br\gamma)$. Even though regions {\it X}-{\it Z} have much
 smaller $\rm EW(Br\gamma)$ values than {\it B}-{\it D}, all these zones are
 consistent with a 6-$\rm 7\,Myr$ stellar population according to models
 (Tables\,\ref{tab:XYZage} and \ref{tab:B_Fage}). In this case, the
 instantaneous model predictions are consistent between our different EW measurements. 

\begin{deluxetable*}{|l|cccc||cc|}\footnote{  The apparent inconsistency between different model predictions for regions
  {\it B}, {\it D} and {\it F} oldest ages is mainly driven by the noisy
  nature of these models for ages $\rm \ge 100\,Myr$.}
\tablecaption{CO$\lambda 2.3\,\mu$m and NaI$\lambda 2.21 \mu$m Instantaneous
  burst model ages for regions B-F\label{tab:B_Fage}} 
\startdata
\hline
\scriptsize{Reg} &  \scriptsize{$\rm Age1_{SB99}^{CO}$} & \scriptsize{$\rm Age1_{\tt STARS}^{CO}$}  &  \scriptsize{$\rm Age2_{SB99}^{CO}$} & \scriptsize{$\rm Age2_{\tt STARS}^{CO}$} &  \scriptsize{$\rm Age1_{\tt STARS}^{NaI}$} &  \scriptsize{$\rm Age2_{\tt STARS}^{NaI}$} \\[0.07cm] 
          &     [Myr]     &        [Myr]  &     [Myr] &     [Myr]     &        [Myr]  &     [Myr]       \\[0.07cm]                             
\hline
{\it B$^{a}$}    &   $7.26_{-0.02}^{+0.02}$   &   $6.72_{-0.06}^{+0.06}$   &   534-592   &    $\ge$ 811    &   $7.89_{-0.05}^{+0.05}$   &        $\ge$ 200    \\[0.07cm]      
{\it C      }    &   $7.33_{-0.02}^{+0.03}$   &   $6.91_{-0.06}^{+0.06}$   &   217-509   &    $\ge$ 200    &   $8.05_{-0.05}^{+0.05}$   &        $\ge$ 200    \\[0.07cm]      
{\it D$^{a}$}    &   $7.30_{-0.03}^{+0.03}$   &   $6.82_{-0.07}^{+0.08}$   &   233-552   &    $\ge$ 811    &   $8.11_{-0.08}^{+0.07}$   &        $\ge$ 200    \\[0.07cm]      
{\it E      }    &   $7.41_{-0.03}^{+0.03}$   &   $7.12_{-0.07}^{+0.07}$   &   200-443   &    $\ge$ 200    &   $8.39_{-0.07}^{+0.07}$   &        $\ge$ 200    \\[0.07cm]      
{\it F$^{a}$}    &   $7.22_{-0.02}^{+0.03}$   &   $6.61_{-0.07}^{+0.08}$   &   581-698   &    $\ge$ 811    &   $8.01_{-0.08}^{+0.08}$   &        $\ge$ 200    \\[0.07cm]      
\enddata
\\[0.24cm] 
\end{deluxetable*}

We have also compared the EW(NaI$\lambda 2.21 \mu$m) measurements with {\tt
  STARS} model predictions. All previous warnings and discussion about the
  EW(CO) age estimation are also applicable to this stellar absorption
  feature. As we can see in Fig.\ref{fig:SSP_models} bottom panel and in
  Tables\,\ref{tab:XYZage} and \ref{tab:B_Fage}, the younger-age predictions
  for an instantaneous model are somehow larger ($\rm \sim 8\,Myr$) than those
  with EW(CO), but still consistent with a young population. 

In summary, while observational evidence points towards two different
stellar populations, the model predictions cannot clearly discriminate
between them. Our measurements are consistent with temporally close,
instantaneous bursts occurred $6$-$\rm 8\,Myr$ ago. Strictly speaking, {\it regions
{\it X}-{\it Z} are always older} than their neighbors {\it B}-{\it D}
(independently of the tracer used) by few hundred thousand years. Even though
it is worth mentioning, such small age differences might be in the limit
of current model capabilities for discriminating between two stellar populations. Certainly, we cannot discard the possibility of two populations of similar ages: a young one ($\rm \sim 7\,Myr$) dominated by O-B stars and an intermediate-age one ($\rm \sim 10\,Myr$) with massive stars which already started their evolution to red supergiants. In such particular case, the presence of local metallicity variations between regions might become an issue given the tight correlation between the CO-index and metallicity (Origlia \& Oliva 2000). In any case, given the above discrepancies, we cannot be conclusive about the existence of one or more stellar populations and thus, further data and tests are necessary to shed light about this issue.

\subsection[]{H$_2$ Excitation Mechanisms}\label{ss:subs25}

Different mechanisms have been proposed to be responsible of the H$_2$
excitation and literature is rich in examples where more than one process
is shaping the observed emission spectra (e.g.\ Draine \& Woods
1990; Mouri 1994; Davies, Sugai \& Ward 1997; Quillen et al.\ 1999; Burston,
Ward \& Davies 2001; Davies et al.\ 2005; Rodr\'iguez-Ardila et al.\ 2005b;
Zuther et al.\ 2007). Between these mechanisms, the most commonly discussed
are UV-fluorescence (photons, non-thermal), shock fronts (collisions, thermal)
and X-ray excitation (collisions through secondary electrons, thermal). 

In this section, we will focus on trying to understand the observed
H$_2$-emission by using different diagnostic diagrams. We will consider both
non-thermal and thermal excitation mechanisms and models in the analysis.

\subsubsection[]{H$_2$ Line Ratios: A Diagnostic Diagram}\label{sss:subs251}  

\begin{figure}
\includegraphics[scale=0.5]{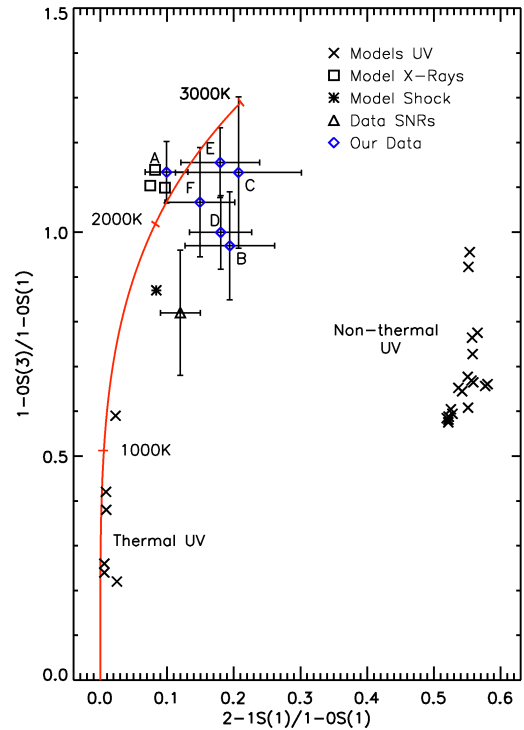}
\caption{\label{fig:test1_H2}\small H$_2$ diagnostic diagram using
  the 2-1\,S(1)/1-0\,S(1) and 1-0\,S(3)/1-0\,S(1) intensity line
  ratios. Diamonds represent our own data for the six selected regions of
  NGC\,5135. The $\times$ symbols represent thermally (Sternberg \& Dalgarno 1989) and non-thermally (Black \& van Dishoeck 1987) UV-exited models. Squares
  correspond to X-ray heating models of Draine \& Woods (1990). The asterisk
  is a shock model from Brand et al.\ (1989). The triangle is SNR data from
  Oliva, Moorwood \& Danziger (1990). The curve represents thermal emission
  between 0 and 3000 K (eqs.\ \ref{eq:boltz} and \ref{eq:coldens}, see references in the text).}
\end{figure}

As a first step, we have calculated different H$_2$-emission line ratios for
our six regions in NGC\,5135. As mentioned in Sec.\,\ref{ss:subs15},
we did not use fluxes corrected for extinction. Different models and data suggest that the excitation mechanisms mentioned above produce characteristic line ratios between different roto-vibrational transitions (see Fig.\,\ref{fig:test1_H2}). In this sense, the many lines presented in our ${\it K}$-band data allowed us to use a line-ratio diagnostic diagram for our analysis. We maximise the difference between thermal and non-thermal processes by trying to remove the ortho/para ratio effect (ratio equal to 3 for a thermal distribution and $< 3$ for a
non-thermal one). We adopt the intensity ratios 2-1\,S(1)/1-0\,S(1) and
1-0\,S(3)/1-0\,S(1) to achieve this goal. Given that the 2-1\,S(1), 1-0\,S(3) and
1-0\,S(1) lines occur in ortho-molecules only, the non-thermal values of their ratios
are supposed to be constant (Mouri 1994). 

In Fig.\,\ref{fig:test1_H2} we present the line-ratio diagnostic diagram for
our six regions compared with different models and data from literature. The
first obvious conclusion from this diagram is that none of our six regions
seem to be dominated by non-thermal (fluorescence) excitation. Instead, they
lie closer to thermal processes like X-ray heating and SNR shocks,
the most likely dominant mechanisms responsible of H$_2$ excitation in the six
zones. Regions {\it B} and {\it D}, however, clearly depart from the pure thermal
emission curve suggesting some UV-fluorescence contribution to the observed
lines. This is reasonable in principle given that {\it B} and {\it D} correspond to recent
star forming regions where UV-photons are abundant. 
The remaining four regions ({\it A}, {\it C}, {\it E} and {\it F}) are
consistent with the purely thermal emission curve (within the uncertainties)
with gas temperatures between 2000 and 3000\,K. The overall lack of dependence on UV-processes (i.\,e.\ recent star formation) agrees with the comparison between our Br$\gamma$ and H2 flux maps (Fig.\,\ref{fig:maps}). The clear structural differences between them implies that the processes exciting the H2 gas are not those who produce Br$\gamma$ (i.\,e.\ recent star formation).

By comparing with models, it is interesting to notice that region {\it A}, the
AGN nucleus, matches very well with the X-ray models of Draine \& Woods
(1990). As mentioned in previous sections, Levenson et al.\ (2004) have
found a strong X-ray source in this area. \footnote{In Levenson et al.\
  (2004) the aperture's
  diameters are about $\rm 4\, \times$ those of our selected regions. We have
  repeated all the tests of this section by using Levenson et al.\ (2004) aperture
  size and no significant changes have been found. The relation between data
  and all model predictions reminds the same, without affecting our
  discussion.} In their discussion, Draine \& Woods (1990, 1991) argued that there is a range of values of X-ray energy per volume for which the gas
gets sufficiently hot to vibrationally excite the H$_2$ (producing the observed
H$_2$ lines with high enough efficiency to be observed), but not hot enough to
dissociate it. They proposed the X-ray emission from SNRs (and not their
shocks) as one plausible stellar energy source. This scenario implies other
constraints like high gas densities ($n_H\ge10^5\,\rm cm^{-3}$, attainable in the
dense circumnuclear regions around AGNs) and high SNe rates (e.g.\, $\rm \approx
1.5\, yr^{-1}$ within the central $\rm 4\,kpc$ of NGC\,6240, but with high
uncertainties). The later constraint seems to be achieved by NGC\,5135, with a SNe rate of $\rm \sim 0.4$-$\rm 0.6\, yr^{-1}$ within the central $\rm 2.3\,kpc$ (see Sec.\,\ref{ss:subs28} for further details). As an extra possible piece of evidence, the Draine \& Woods' models predict a decrement in the 2-1\,S(3) transition flux ($v=2$, $J=5$ level), which we actually observed in Fig.\,\ref{fig:test2_H2} (within large uncertainties).

We also highlight in Fig.\,\ref{fig:test1_H2} that region {\it E} is consistent
with a thermal process but far from the SNR data and shock model. As we saw in Sec.\,\ref{ss:subs24}, SNR shocks were an important candidates for H$_2$-excitation given the high [FeII] fluxes and FWHMs observed in this area. Could region {\it E} be in a similar scenario to the one described above for {\it A}? Certainly, the X-ray models are not as close to {\it E} as to region {\it A}, however we know from Levenson et al.\ (2004) work that the second X-ray peak comes from region {\it E} and it has a
star formation origin. In principle, the same arguments presented before for
region {\it A} might be applied to {\it E} in the sense that an important
fraction of H$_2$ excitation might be produced by X-rays of stellar
origin. 

Finally, region {\it F} is also consistent with thermal
processes. However, the uncertainties make it difficult to distinguish between
X-ray and SNR shock scenarios. Because the later has a dominant role in
ionizing the local ISM gas (Sec.\,\ref{ss:subs241}), SNR shocks become a
likely candidate to dominate the H$_2$ local excitation in this area.

\subsubsection[]{H$_2$ population diagram}\label{sss:subs252}  

Because we are able to measure H$_2$-emission lines from different vibrational
transitions (1-0 and 2-1), we adopt another approach to characterise the
observed excitation mechanisms. We have just seen in Sec.\,\ref{sss:subs251}
that the dominant processes responsible for H$_2$ excitation are
thermal (collisional). Therefore, if we assume pure thermal excitation, the different
roto-vibrational levels are populated as described in the Boltzmann
equation. This allowed us to represent the different transitions as a function of
their population (probability) densities versus the energy of the upper level,
in what is called a population diagram. For a
thermally excited gas, all the transition values lie on a straight line and the
corresponding slope is inversely proportional to the H$_2$ gas temperature.

Given that molecular gas clouds contain a large number of molecules, the
population densities are essentially proportional to the observed column
densities. So, the probability ratio between two different states can be
expressed with the Boltzmann equation as
\begin{equation}\label{eq:boltz}
\frac{P_a}{P_b} = \frac{N(v,J)_a}{N(v,J)_b} = \frac{g_a\cdot e^{-E_a/kT}}{g_b\cdot e^{-E_b/kT}}\end{equation}
where $P_a$ and $P_b$ are the probabilities of finding an H-molecule in any of
the degenerate states ($g_a$, $g_b$) with energy $E_a$, $E_b$. $N(v,J)_a$ and
$N(v,J)_b$ are the observed column densities at the corresponding states, $v$
and $J$ are the vibrational and rotational quantum numbers, respectively. $k$ is the Boltzmann constant and $T$ is the gas
temperature. The column densities can be derived with the formula:
\begin{equation}\label{eq:coldens}
N(v,J) = \frac{f}{A(v\,J, v^{\prime}\,J^{\prime})} \times \frac{\lambda}{hc} \times \frac{4 \pi}{\Omega_{aper}}
\end{equation}
where $f$ is the measured flux for the corresponding emission line, $A(v\,J,
v^{\prime}\,J^{\prime})$ is the transition probability from the $(v\,J)$
to the $(v^{\prime}\,J^{\prime})$ quantum state (taken from Wolniewicz et
al.\ 1998), $\lambda$ is the rest frame
line wavelength, $h$ is the Planck constant, $c$ is the speed of light and
$\Omega_{aper}$ is the squared aperture diameter of
$\rm 0.^{\prime \prime}62$. The derived column densities for each H$_2$
indicator and region are presented in Table\,\ref{tab:colden}.

\begin{table*}  
\begin{center}
 \caption{H$_2$ column densities derived for six regions in NGC~5135.\label{tab:colden}} 
 \begin{tabular}{@{}|l|cc|cc|cc|cc|cc|cc|@{}}
 \hline
 \hline
         &  Region {\it A} & & Region {\it B} & & Region {\it C} & & Region {\it D} & & Region {\it E} & & Region {\it F} & \\
   Line  & $N_{H, obs}^a$ &  $N_{H, cor}^b$  & $N_{H, obs}^a$ &  $N_{H, cor}^b$  & $N_{H, obs}^a$ &  $N_{H, cor}^b$  & $N_{H, obs}^a$ &  $N_{H, cor}^b$  & $N_{H, obs}^a$ &  $N_{H, cor}^b$  & $N_{H, obs}^a$ &  $N_{H, cor}^b$    \\
         &   \tiny{[$\rm 10^{21} cm^{-2}$]}  &  & \tiny{[$\rm 10^{21} cm^{-2}$]} &    &   \tiny{[$\rm 10^{21} cm^{-2}$]}  &   & \tiny{[$\rm 10^{21} cm^{-2}$]} &    &   \tiny{[$\rm 10^{21} cm^{-2}$]}  &  & \tiny{[$\rm 10^{21} cm^{-2}$]} &   \\
 \hline
 \hline
  1-0\,S(0)   & 3.49      & 9.35       &  1.11       &1.87       & 0.88       & 1.60      & 1.57       & 2.36      & 1.93      & 3.13      & 0.85      & 1.51\\    
              &$\pm$0.66  & $\pm$ 2.17 &  $\pm$0.26  &$\pm$0.49  & $\pm$0.28  & $\pm$0.56 & $\pm$0.30  & $\pm$0.54 & $\pm$0.38 & $\pm$0.73 & $\pm$0.20 & $\pm$0.41\\ 
  1-0\,S(1)   &10.54      &30.38       &  2.78       &4.85       & 2.45       & 4.68      & 4.51       & 6.97      & 6.94      &11.65      & 2.37      & 4.38\\ 
              &$\pm$0.46  & $\pm$4.64  &  $\pm$0.22  &$\pm$0.78  & $\pm$0.25  & $\pm$0.81 & $\pm$0.24  & $\pm$1.05 & $\pm$0.33 & $\pm$1.73 & $\pm$0.18 & $\pm$0.73\\ 
  1-0\,S(2)   & 3.16      & 9.78       &  0.90       &1.63       & 1.07       & 2.13      & 1.33       & 2.12      & 2.41      & 4.20      & 0.83      & 1.60\\ 
              &$\pm$0.40  &$\pm$ 1.96  &  $\pm$0.21  &$\pm$0.45  & $\pm$0.27  & $\pm$0.63 & $\pm$0.21  & $\pm$0.46 & $\pm$0.30 & $\pm$0.82 & $\pm$0.17 & $\pm$0.41\\ 
  1-0\,S(3)   & 9.10      &30.20       &  2.05       &3.86       & 2.11       & 4.40      & 3.43       & 5.62      & 6.11      &10.99      & 1.93      & 3.86\\ 
              &$\pm$0.38  &$\pm$5.17   &  $\pm$0.19  &$\pm$0.71  & $\pm$0.23  & $\pm$0.85 & $\pm$0.21  & $\pm$0.96 & $\pm$0.29 & $\pm$1.82 & $\pm$0.16 & $\pm$0.72\\ 
  2-1\,S(1)   & 0.77      & 2.04       &  0.40       &0.66       & 0.38       & 0.68      & 0.60       & 0.90      & 0.92      & 1.48      & 0.26      & 0.46\\ 
              &$\pm$0.25  &$\pm$ 0.70  &  $\pm$0.13  &$\pm$0.24  & $\pm$0.17  & $\pm$0.31 & $\pm$0.15  & $\pm$0.25 & $\pm$0.30 & $\pm$0.52 & $\pm$0.09 & $\pm$0.17\\ 
  2-1\,S(3)   & 0.54      & 1.64       &  0.25       &0.44       & 0.15       & 0.29      & 0.26       & 0.42      & 0.38      & 0.64      & 0.16      & 0.29\\ 
              &$\pm$0.22  &$\pm$ 0.70  &  $\pm$0.11  &$\pm$0.21  & $\pm$0.12  & $\pm$0.24 & $\pm$0.10  & $\pm$0.17 & $\pm$0.15 & $\pm$0.27 & $\pm$0.08 & $\pm$0.16\\          
\hline
\end{tabular}
\end{center}
\footnotesize{($^a$) Total column densities derived by using the observed H$_2$
  line fluxes. Below the main values, their $\pm 1\,\sigma$ errors.\\ ($^b$) Total column densities after extinction correction for
  each observed H$_2$ line and region as listed in
  Table\,\ref{tab:lineparam}. Below the main values, their $\pm 1\,\sigma$ errors.}
\end{table*}

By taking the logarithm in eq.\,\ref{eq:boltz} we end up with a generic
expression for $N(v,J)$ of the form
\begin{equation}\label{eq:popdiag}
ln \bigg( \frac{N(v,J)/g_J}{N(1,3)/g_3} \bigg) = \frac{-E(v,J)/k}{T}+Constant
\end{equation}
where we have chosen the transition $(v,J)=(1,3)$ (corresponding to the
1-0\,S(1) line) to normalise the derived column densities. The constant is
independent of the transition. Because the ratios $N(v,J)/N(1,3)$ are
equivalent to flux ratios, we have used the none extinction corrected fluxes
for the population diagram as argued in Sec.\,\ref{ss:subs15}. In any case, we
present in Table\,\ref{tab:colden} the column densities derived from each
H$_2$ line with and without extinction correction. 

In general, our corrected column densities, $N_{H,cor}$, range between $\sim
10^{21}$-$10^{22}\rm \,cm^{-2}$ for all indicators and regions, with slightly
lower densities derived from the $v =$ 2-1 transitions. The larger $N_{H,corr}$ values coming from region {\it A} ($ 1$-$3\times 10^{22}\rm \,cm^{-2}$) are not as large as those derived by Levenson et al.\ (2004) ($>10^{24}\rm \,cm^{-2}$, from X-ray observations). Gonz\'alez Delgado et al.\ (1998, using UV data) and Levenson et al.\ (2004) have also reported column densities for {\it E}
($9.9\times 10^{20}$ and $1.1\times 10^{21}\rm \,cm^{-2}$, respectively but
within larger apertures than ours) which are consistent with many of our
estimations in this area.

\begin{figure}
\includegraphics[scale=0.46]{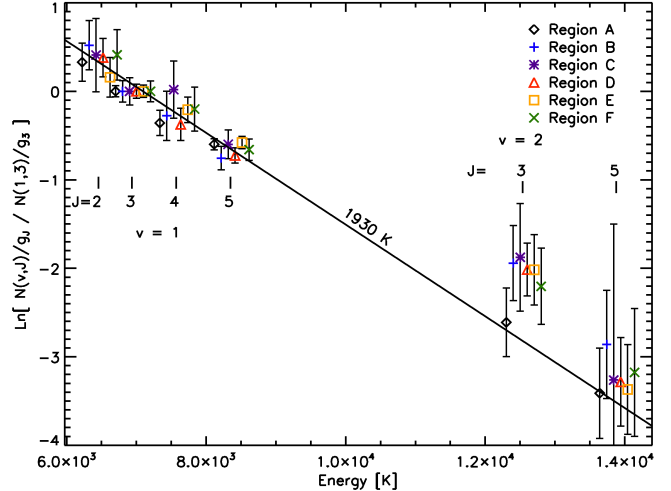}
\caption{\label{fig:test2_H2}\small H$_2$ level population diagram relative to
  1-0S(1) for the six regions of NGC\,5135. The data points are slightly spread
  in the X-direction simply for illustrative purposes. The straight line is the linear
  fit to the $v =$ 1-0 levels. Its slope is consistent with an isothermal
  population at $1930 \pm 114$\,K.}
\end{figure}

In Fig.\,\ref{fig:test2_H2} the H$_2$ population diagram for our six regions
is presented, showing the column densities normalised by
the (1,3) transition versus the energy of the upper level (in Kelvin). We have
used six different emission lines including four $v =$ 1-0 transitions (S(0),
S(1), S(2) and S(3)) and two $v =$ 2-1 transitions (S(1) and S(3)). The
different symbols represent our six regions of integrated spectra.

The $v =$ 1-0 transitions are
clearly well thermalized in all regions, consistent with a similar thermal excitation temperature. By a linear
least square minimisation to the $v =$ 1-0 transitions, we have found a
common excitation temperature of $1930 \pm 114$\,K which is in agreement with
the temperature range $2000-3000$ K derived from Fig.\,\ref{fig:test1_H2}. The
reason for using only the $v =$ 1-0 lines for the fit resides in that these
lower energy transitions are mainly susceptible to thermal processes, while
higher energy transitions ($v =$ 2-1) are more easily affected by fluorescence
and other non-thermal mechanisms (if present).

For the $v =$ 2-1 transitions, however, the situation looks dissimilar. While
the (2,5) transitions are consistent with the thermal linear fit, the (2,3)
ones are shifted upwards this relation. The exception is region {\it A} for
which all transitions are consistent with a thermal mechanism. The (2,3)
deviation of the remaining areas could be explained by some contribution from
UV-fluorescence, expected to be present in starforming regions. However, when
fluorescence dominates the H$_2$ excitation, a larger positive offset between
the 2-1 transitions and our fit would be expected, with line ratios like
2-1S(1)/1-0S(1) of about 0.5. The observed values in Fig.\,\ref{fig:test1_H2}
are rather smaller ($<\,0.2$), suggesting that UV-fluorescence might be
present but does not dominates the H$_2$ excitation. The (2,5) transition case
is rather different, with larger uncertainties in the data. However we just
point out that almost all data points lie very close to the thermal linear fit,
which is not expected for $v =$ 2-1 transitions if fluorescence is present. As
mentioned in Sec.\,\ref{sss:subs251}, the Draine \& Woods (1990) X-ray models
predict a (2,5) transition decrement which may explain the observed behaviour
but, in principle, only for regions {\it A} and {\it E} (the only two with
clear X-ray sources). The existence of a
second thermal component for the $v =$ 2-1 transitions cannot be explored
given the uncertainties.

\subsection[]{[FeII] as a SNe tracer}\label{ss:subs28}
For many years different authors have suggested that the [FeII]-emission from
galaxies traces the fast shocks produced by SNR and so, their SNe activity
(e.g.\ Graham et al.\ 1987, Moorwood \& Oliva 1988, Oliva et al.\ 1989,
Greenhouse et al.\ 1991, Colina 1993, Alonso-Herrero et al.\ 2003, Labrie \& Pritchet 2006). The partial ISM ionization produced by this mechanism is
suitable for most of the Fe atoms to be in low ionization states.

Different authors have derived empirical correlations we can use to
estimate the SNe rate directly from the [FeII] data. Calzetti (1997) has found
the following expression for [FeII] measurements
\begin{equation}\label{eq:snr_calzetti}
SN_{rate, Ca97}(yr^{-1})= 5.38\cdot L_{[FeII]}(W) \times 10^{-35},
\end{equation}
for which SNes release $\rm 10^{51}\,erg$ of energy during their lives. Also,
from their study of SNR in M82 and NGC\,253, Alonso-Herrero et al.\ (2003)
have found that  
\begin{equation}\label{eq:snr_alonso}
SN_{rate, AH03}(yr^{-1})= 8.0\cdot L_{[FeII]}(W) \times 10^{-35}.
\end{equation}

In Table\,\ref{tab:sne_rate} we present the derived SNe rates for regions {\it
  B}-{\it F} by using eqs.\,\ref{eq:snr_calzetti}-\ref{eq:snr_alonso}. Region
{\it A} has been excluded from the analysis because it is likely that the
  [FeII]-emission comes from the AGN and not from SNe. 

\begin{table}  
\begin{center}
 \caption{SNe Rates derived from [FeII] emission\label{tab:sne_rate}} 
 \begin{tabular}{@{}|l|cc|@{}}
\hline
 \scriptsize{Reg}        &  Ca97 & A-H03 \\
                         & $\rm [yr^{-1}]$   & $\rm [yr^{-1}]$\\
\hline
{\it B}$^b$    &  0.006 $\pm$ 0.001 &  0.009 $\pm$ 0.002   \\
{\it C}$^b$    &  0.006 $\pm$ 0.001 &  0.009 $\pm$ 0.002   \\
{\it D}$^b$    &  0.011 $\pm$ 0.002 &  0.017 $\pm$ 0.004   \\
{\it E}$^b$    &  0.027 $\pm$ 0.006 &  0.041 $\pm$ 0.009   \\
{\it F}$^b$    &  0.006 $\pm$ 0.001 &  0.009 $\pm$ 0.002   \\
\hline
Tot.\,-{\it A}$^{a,b}$    &  0.368 $\pm$ 0.084 &  0.547 $\pm$ 0.125   \\
\hline
\end{tabular}\\
\end{center}
\footnotesize{($^a$) The complete {\it H}-band FoV was collapsed in a single
  spectrum and the measured [FeII] flux was extinction corrected by our median
  FoV-extinction value of $A_V=6\rm \,mag$. For this calculation, the [FeII]
  flux from region {\it A} has been subtracted from the total because it
  likely comes from the AGN and not from SNe. ($^b$) Our extinction corrected
  [FeII] fluxes were used. For {\it B}-{\it F} it corresponds to apertures of
  157\,pc in diameter.}
\end{table}

We remind the reader that all SNe rates derived from [FeII] are {\it upper limits}
because we are assuming that {\it all} [FeII] flux comes from SNR. We have
already seen in previous sections that there exist other processes that might
contribute to [FeII] enhancement.

The total-FoV SNe rate from Table\,\ref{tab:sne_rate} is rather high compared to other well studied starburst galaxies. The rates derived for the entire galaxies M82 and NGC\,253 by Alonso-Hererro et al.\ (2003) of $\rm \approx 0.1\, yr^{-1}$ are about 5$\,\times$ lower than our [FeII] estimation between the central
$\rm 2.3\,kpc \times 2.3\,kpc$ of NGC\,5135. The source of this
discrepancy is the different SFRs of these galaxies. By using Sanders et
al.\,(2003) {\tt IRAS} data and the Kennicutt (1998) law, we estimate total SFRs
of 15.0, 3.4 and $\rm 7.0\,M_{\odot} yr^{-1}$ for NGC\,5135, NGC\,253 and M82,
respectively. For NGC\,5135, we subtract the AGN contribution (25\%,
Alonso-Herrero et al.\,2006b) from the total far-IR luminosity used in this
calculation. As we can see, the different SNe rates between NGC\,5135 and
NGC\,253 can be explained by the stronger SFR of the
former. The relative SFRs of NGC\,5135 and M82 can also explain their
different SNe rates between a factor of 2. Such a difference can be easily
attributed to the many uncertainties of this calculation, including the flux
measurements, the presence of different alternative sources for those fluxes
and the implicit models assumptions for the SNe rates and SFRs determination.
\\

\begin{figure}
\includegraphics[scale=0.48]{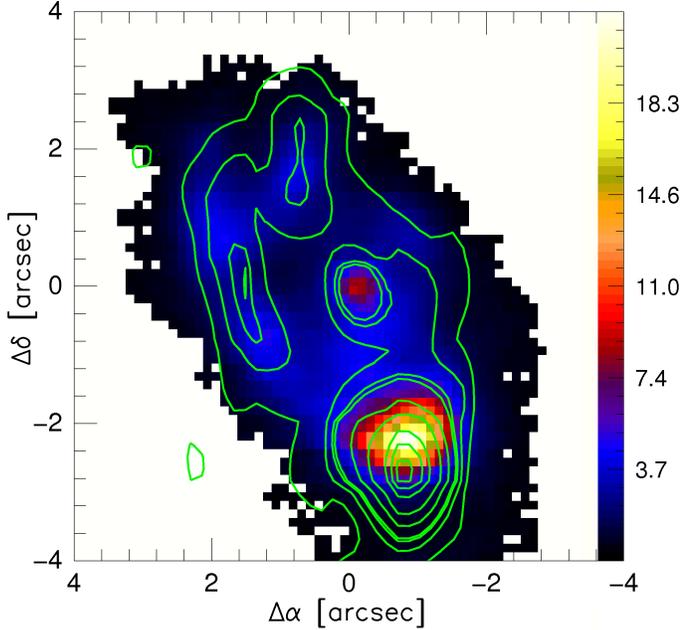}
\caption{\label{fig:feII_radio}\small [FeII] flux map. The
  contours are the $\rm 6\,cm$ VLA radio emission from Ulvestad \& Wilson (1989).}
\end{figure}

Forbes \& Ward (1993) have found a tight correlation between the 6\,cm and
[FeII]-emission and it is also well known that SNe and SNR shells are strong radio emitters. Therefore, to first order, a spatial correspondence between radio
and [FeII]-emission could be expected. To test this, in
Fig.\,\ref{fig:feII_radio} we compare our [FeII] flux map with VLA $\rm 6\,cm$
radio emission (contours, $\rm 0.91^{\prime \prime} \times 0.60^{\prime
  \prime}$ beam resolution) obtained by Ulvestad \& Wilson (1989). Both
images were aligned by matching the corresponding nuclear ({\it A} region)
unresolved peak. As we can see, the spatial structures are in good
agreement, suggesting that at least part of the [FeII]-emission comes from
SNRs.

We can go further on our comparison by contrasting the SNe rates derived from
radio and [FeII]-emission in our {\it B}-{\it F} regions. To do so, however, we have the limitation of the poorer spatial resolution
of the 6\,cm VLA image. Therefore, we decided to keep the centers of our {\it
  B}-{\it F} regions but use a larger circular aperture of $\rm 0.91^{\prime \prime}$
($\rm \approx 260\,pc$ in diameter, the larger dimension of the radio beam
size). We label our $\rm 0.91^{\prime \prime}$ aperture regions
as {\it B$^{\prime}$}-{\it F$^{\prime}$}. In this way, both our [FeII] and 6\,cm emission
were remeasured within the new apertures. We highlight that, for this larger
aperture, regions {\it D$^{\prime}$} and {\it E$^{\prime}$} share about
40\,$\%$ of their spaxels.

In a similar way to [FeII], other works have explored the
relation between the 6\,cm radio flux and the SNe rate. By using models and
empirical correlations, the different authors have made diverse assumptions
concerning the origin of their radio observations. From their study of compact radio
sources in M82, Huang et al.\ (1994) derived the following expression

\begin{equation}\label{eq:snr_huang}
SN_{rate, Hu94}^{6\rm cm}(yr^{-1})= 0.22 \cdot L_{\nu}(W\,Hz^{-1}) \times 10^{-22}.
\end{equation}
These authors assume that the entire non-thermal radio emission of M82 comes
from SNRs. $L_{\nu}$ is the 6\,cm radio flux in [$\rm W\,Hz^{-1}$].    

In a study of the brightest radio sources in Arp\,299, Neff et al.\ (2004)
derived a similar expression for the SNe rate as a function
of the 6\,cm radio flux from SNRs
\begin{displaymath}
SN_{rate, Ne04}^{6\rm cm}(yr^{-1})=  
\end{displaymath}
\begin{equation}\label{eq:snr_neef}
\frac{0.37\cdot f \cdot L_{\nu}(W\,Hz^{-1})\cdot 10^{33}}{5\,{\rm mJy}\cdot 4\pi(D_{Arp299})^2}\times \bigg(\frac{408}{4885}\bigg)^{\alpha+0.75}
\end{equation}

According to the authors, this equation is valid for few hundred parsec
spatial scales, the same order of our own aperture diameters. Eq.\,\ref{eq:snr_neef} has been modified to adapt the measured radio flux in NGC\,5135 to the distance of Arp\,299 ($\approx$ 42\,Mpc, for which the original expression was made). The spectral index $\alpha$ was chosen to be $-0.58$, a value consistent with typical Galactic SNRs (Weiler et al.\ 2002) and with the Neff et al.\ measurements in Arp\,299. The
parameter $f$ is the fraction of radio flux density that can be associated
with individual SNRs on parsec scales. Because there exist other sources of
radio emission apart of SNR, we adopt a $f$ value of 0.15
following Neff et al. An $f$ value of 0.1-0.2 provides a calibration similar to
those obtained by Condon (1992) and Huang et al.\ (1994) works on radio
sources in starburst galaxies. Also, the chosen value of $f$ is in excellent
agreement with Colina \& P\'erez-Olea (1995) models: for the majority of them
the fraction of SNR radio emission with respect to other sources (free-free
emission, young radio SNe) ranges between 9-19\% of the total flux. 

Finally, we have also used P\'erez-Olea \& Colina (1995) model predictions for
the SNe rate, based on our $6\rm cm$ measurements and Br$\gamma$ age
estimations. An instantaneous burst and Salpeter IMF have been assumed, and
different masses M$_{up}$ (30, 60, 1200$\,\rm M_{\odot}$) and M$_{low}$ (0.85,
3$\,\rm M_{\odot}$) have been considered for the calculations. These models
make a detailed consideration of the different radio components, including
free-free radiation, young radio SNe and older SNRs. In Table\,\ref{tab:sne_rate_prime} we present the SNe rates derived from [FeII] and 6\,cm radio emission (eqs.\,\ref{eq:snr_calzetti}-\ref{eq:snr_neef}) for regions {\it B$^{\prime}$}-{\it F$^{\prime}$}.

\begin{table*}  
\begin{center}
 \caption{SNe Rates derived from [FeII] and Radio emission \label{tab:sne_rate_prime}} 
 \begin{tabular}{@{}|l|cc||ccc|@{}}
\hline
 \scriptsize{Reg$^a$}        &  \scriptsize{Ca97$_{\rm [FeII]}$} & \scriptsize{A-H03$_{\rm [FeII]}$} & \scriptsize{Hu94$_{\rm 6cm}$} &
  \scriptsize{Ne04$_{\rm 6cm}$} & \scriptsize{P-O95$_{\rm 6cm}^b$}\\
                         & $\rm [yr^{-1}]$   & $\rm [yr^{-1}]$ & $\rm [yr^{-1}]$ & $\rm [yr^{-1}]$ & $\rm [yr^{-1}]$\\
\hline
{\it B$^{\prime}$}     & 0.012 $\pm$ 0.002&  0.017 $\pm$ 0.004 & 0.012 $\pm$ 0.004  & 0.019 $\pm$ 0.007 & 0.013-0.014 \\[0.1cm]
{\it C$^{\prime}$}     & 0.012 $\pm$ 0.003&  0.018 $\pm$ 0.004 & 0.010 $\pm$ 0.005  & 0.016 $\pm$ 0.008 & 0.011-0.012 \\[0.1cm]
{\it D$^{\prime}$}     & 0.024 $\pm$ 0.005&  0.036 $\pm$ 0.008 & 0.032 $\pm$ 0.075  & 0.051 $\pm$ 0.118 & 0.036-0.037 \\[0.1cm]
{\it E$^{\prime}$}     & 0.051 $\pm$ 0.012&  0.076 $\pm$ 0.016 & 0.064 $\pm$ 0.001  & 0.100 $\pm$ 0.002 & 0.071-0.074 \\[0.1cm]
{\it F$^{\prime}$}     & 0.011 $\pm$ 0.002&  0.016 $\pm$ 0.004 & 0.012 $\pm$ 0.004  & 0.019 $\pm$ 0.006 & 0.018-0.019 \\[0.1cm]
\hline
\end{tabular}\\
\end{center}
\footnotesize{($^a$) Same centers as {\it B}-{\it F} regions with aperture of $\rm 0.91^{\prime \prime}$.}
\footnotesize{($^b$) For P\'erez-Olea \& Colina (1995) models, we present ranges of SNe
  rates for each region, where different upper- and lower-limit
  masses have been considered for a Salpeter IMF.}
\end{table*}

Table\,\ref{tab:sne_rate_prime} clearly shows the excellent agreement between [FeII] and radio SNe rate predictions for these $\rm \sim 200\,pc$-scale regions. This agreement is totally independent of the correlation or model used to estimate the SNe rate. These results provide additional support to previous findings in normal starburst galaxies like M82 and NGC\,253 where similar SNe rates are predicted from [FeII] and $\rm 6\,cm$ radio emission. The radio predictions in this table are of the same order as those of the brightest radio region in Arp\,299 ($0.5$-$1.0\, \rm yr^{-1}$, Neff et al.\ 2004), a bright IR luminous galaxy.

\section{Summary}\label{s:conclu}

In this paper we have studied the central $\rm 2.3\,kpc$ of
NGC\,5135, a nearby LIRG galaxy. By using new NIR IFU spectroscopy
(VLT/SINFONI) we have traced different structural components including the
coronal gas (an AGN-activity tracer), ionized (young stellar population) and
partially ionized gas (SNe remnants), warm molecular gas and the
giant/supergiant stellar population. We have selected individual regions of
interest for a detailed analysis (see Fig.\,\ref{fig:maps} top panels, and
Fig.\,\ref{fig:stelmaps} top-left panel, for their spatial identification). By
using these data, we have shown the potential of the IFS technic when applied to a prototype LIRG: studying the complex spatial structure of its different components, the different gas ionization/excitation mechanisms, the AGN activity, the SNe rates, the internal gas/dust
extinction and the stellar population ages. Our conclusions are:
\\

1) Overall, we have found that SNR shocks play a dominant role on ionizing/exciting the different gas phases in the central $\rm 2.3\,kpc$ of NGC\,5135. Our NIR data reveals that other processes associated to recent star formation and AGN activity usually have a rather localised and smaller
overall impact. By using Br$\gamma$-emission, we estimate the AGN contribution to
the gas ionization in $\lesssim 3\%$ of the total within our FoV. In the same area, the AGN contributes in $\lesssim 3$-$5\%$ to the total NIR luminosity.
\\

2) We have found the largest ($\rm \sim 600\,pc$ in projection)
[SiVI]-ionization cone reported in literature until now. This structure is at
least $4 \times$ larger than any previous detection in active
galaxies. Pointing in opposite direction, a fainter counter-[SiVI]-cone has been also
detected. 
\\

3) The SNe rates derived from the [FeII] emission are in excellent agreement with
$\rm 6\,cm$ radio emission predictions, reinforcing the use of [FeII] as a SNe
activity tracer. Typical rates between $0.01$-$\rm 0.04\,yr^{-1}$ were found for
individual $\rm \sim 200\,pc$ regions. The [FeII] prediction for our entire
FoV ($0.4$-$0.5\rm\,yr^{-1}$) is rather high compared to other starburst
galaxies. Such a difference, however, can be explained by the higher SFR of
NGC\,5135. 
\\

4) In a complex spatial distribution, different processes drive the gas
ionization depending on the region of interest. These include photoionization
from starforming knots ($\HII$ regions {\it B}, {\it C} and {\it D}), SNe remnant
shocks ({\it E} and {\it F}) and AGN photoionization (region {\it
  A}). Mechanisms like stellar X-rays photoionization and star formation might
also be present in other regions ({\it E} and {\it A}, respectively). 
\\

5) The molecular gas excitation mechanisms are mainly thermal ($\rm \sim
1930\pm 114\,K$) in all galaxy regions, being X-rays and SNe remnant shocks
the dominant processes. Contrary to the expectations we might have for a
galaxy with recent star formation, UV-processes like fluorescence have, at the
most, a rather marginal contribution even in $\HII$ regions ({\it B}, {\it C}
and {\it D}). The column densities found ($\rm N_H\sim 10^{21}$-$\rm
10^{22}\,cm^{-2}$) are consistent with previous literature estimations at other
wavelengths.
\\ 

6) As upper limit estimations, the last burst of star formation occurs about
$\rm 6$-$\rm 8\,Myr$ ago in all regions, consistent with previous findings in
the UV/optical.  The presence of a second, older stellar population dominated by
bright red giant/supergiant stars is strongly suggested by the data. Although
NIR-continuum, CO$\lambda2.29\mu$m and NaI$\lambda 2.21 \mu$m observations
agree with this idea, the simple stellar population models cannot clearly
discriminate between both populations.
\\

7) For the nucleus (region {\it A}), the presence of deep CO$\lambda2.29\mu$m and
NaI$\lambda 2.21 \mu$m features highlight the importance of star formation in
the AGN neighbourhood ($\rm \leq 90\,pc$ from the center). This is also
consistent with previous findings in X-rays.
\\

8) A mean internal visual extinction of $A_V=\rm 6\,mag$ ($\rm \approx 0.55\,mag$ at
2.2\,$\mu$m) has been measured in the central $\rm 2.3\,kpc$ of this galaxy. For the nucleus (region {\it A}) a higher mean value of $A_V=\rm 10\,mag$ was found, with a peak of $\rm 17\,mag$ in the very center of this area. This result is qualitatively consistent with high extinctions reported for this region in the literature.

\section*{Acknowledgements}
The authors thank Dr.\,M.\,Garc\'ia-Mar\'in for her support during phase~II
 observation preparation. We also thank Dr. R.\,I.\,Davies, A.\,Sternberg,
 J.\,Falc\'on-Barroso, L.\,Origlia, A.\,Labiano, J.\,Rodr\'iguez-Zaur\'in and
 T.\,D\'iaz-Santos for their help and fruitful discussion. A.\,G.\,B.\, thanks
 the financial support of Comunidad Aut\'onoma de Madrid through the ASTRID
 program. 

Based on observations carried out at the European Southern Observatory,
Paranal (Chile), program 077.B-0151(A). This work has been supported by the
Spanish Ministry for Science and Innovation under grant ESP2007-65475-C02-01.

\clearpage
\begin{landscape}
\begin{deluxetable}{|lc|cccc|cccc|}
\tablecaption{Fluxes, FWHM and EW for six regions in NGC\,5135 \label{tab:lineparam}} 
\startdata
\hline
        &  &  Region\,{\it A} &\scriptsize{$\langle A_V\rangle=9.71\pm 1.34$} & & &  Region\,{\it B} &\scriptsize{$\langle A_V\rangle=5.11\pm 1.29$} & &   \\[0.07cm]
\hline
Line & $\lambda_c$ &Flux\tablenotemark{a} & Corr.Flux\tablenotemark{b} & FWHM& EW & Flux\tablenotemark{a} & Corr.Flux\tablenotemark{b} & FWHM& EW  \\[0.07cm]
     & $\rm [\mu m]$& \scriptsize{$[\rm erg\,\,s^{-1}\,cm^{-2}]\times 10^{-16}$}& \scriptsize{$[\rm erg\,\,s^{-1}\,cm^{-2}]\times 10^{-16}$}& [$\rm \AA$]& [$\rm \AA$] & \scriptsize{$[\rm erg\,\,s^{-1}\,cm^{-2}]\times 10^{-16}$}& \scriptsize{$[\rm erg\,\,s^{-1}\,cm^{-2}]\times 10^{-16}$}& [$\rm \AA$]& [$\rm \AA$]\\
\hline 
1-0\,S(0)   &2.22 &  2.69 $\pm$ 0.50 & 7.20 $\pm$ 1.67 &   14.41 $\pm$ 2.09 &   2.08 $\pm$ 0.39   & 0.86 $\pm$ 0.20&  1.44 $\pm$ 0.38 &   8.70 $\pm$ 1.48  &  2.47 $\pm$ 0.57  \\[0.07cm]  
1-0\,S(1)   &2.12 & 11.66 $\pm$ 0.51 &33.62 $\pm$ 5.14 &   13.06 $\pm$ 0.41 &   9.28 $\pm$ 0.40   & 3.07 $\pm$ 0.25&  5.36 $\pm$ 0.87 &  10.64 $\pm$ 0.61  &  8.54 $\pm$ 0.69  \\[0.07cm]  
1-0\,S(2)   &2.03 &  4.19 $\pm$ 0.52 &12.97 $\pm$ 2.60 &   14.11 $\pm$ 1.30 &   3.34 $\pm$ 0.42   & 1.20 $\pm$ 0.27&  2.17 $\pm$ 0.59 &  12.33 $\pm$ 2.09  &  3.10 $\pm$ 0.71  \\[0.07cm]  
1-0\,S(3)   &1.96 & 13.22 $\pm$ 0.56 &43.86 $\pm$ 7.51 &   15.03 $\pm$ 0.46 &  10.64 $\pm$ 0.45   & 2.98 $\pm$ 0.28&  5.60 $\pm$ 1.04 &  12.22 $\pm$ 0.84  &  7.25 $\pm$ 0.68  \\[0.07cm]  
2-1\,S(1)   &2.25 &  1.16 $\pm$ 0.37 & 3.05 $\pm$ 1.06 &    9.19 $\pm$ 2.20 &   0.90 $\pm$ 0.29   & 0.60 $\pm$ 0.20&  0.99 $\pm$ 0.36 &   9.83 $\pm$ 2.47  &  1.77 $\pm$ 0.60  \\[0.07cm]  
2-1\,S(3)   &2.07 &  1.03 $\pm$ 0.41 & 3.08 $\pm$ 1.31 &   11.04 $\pm$ 3.30 &   0.81 $\pm$ 0.32   & 0.47 $\pm$ 0.21&  0.84 $\pm$ 0.40 &   9.55 $\pm$ 3.23  &  1.24 $\pm$ 0.56  \\[0.07cm]  
Br$\gamma$  &2.16 &  5.18 $\pm$ 0.57 &14.46 $\pm$ 2.60 &   21.48 $\pm$ 1.76 &   4.05 $\pm$ 0.45   & 7.26 $\pm$ 0.29& 12.45 $\pm$ 1.77 &  12.00 $\pm$ 0.33  & 20.54 $\pm$ 0.83  \\[0.07cm]    
Br$\delta$  &1.94 &  2.65 $\pm$ 0.51 & 8.89 $\pm$ 2.27 &   11.85 $\pm$ 1.68 &   2.11 $\pm$ 0.41   & 3.94 $\pm$ 0.32&  7.44 $\pm$ 1.34 &   9.91 $\pm$ 0.56  &  9.42 $\pm$ 0.76  \\[0.07cm]     
$\rm [SiVI]$&1.96 & 26.83 $\pm$ 0.80 &88.53 $\pm$14.86 &   27.95 $\pm$ 0.59 &  20.78 $\pm$ 0.62   & ---       ---  &  ---       ---   &   ---  ---         &  ---        ---   \\[0.07cm]    
$\rm [FeII]$&1.64 & 17.04 $\pm$ 1.20 &82.30 $\pm$18.86 &   15.76 $\pm$ 0.81 &   6.49 $\pm$ 0.46   &11.54 $\pm$ 0.71& 26.44 $\pm$ 5.76 &   10.16 $\pm$ 0.44  & 10.48 $\pm$ 0.64  \\[0.07cm] 
HeI         &2.06 &  2.74 $\pm$ 0.49 & 8.30 $\pm$ 1.96 &   14.83 $\pm$ 1.98 &   2.16 $\pm$ 0.39   & 1.94 $\pm$ 0.26&  3.48 $\pm$ 0.69 &   11.70 $\pm$ 1.13 &   5.08 $\pm$ 0.67 \\[0.07cm]  
CO\tablenotemark{c} &2.29 & 6.76 $\pm$ 0.18 & 7.37 $\pm$ 1.61 &   ---   ---        &   2.15 $\pm$ 0.06   & 3.96 $\pm$ 0.23& 4.31 $\pm$ 0.97 &   ---  ---          & 4.98 $\pm$ 0.29  \\[0.07cm] 
NaI\tablenotemark{c}&2.21 &  2.36 $\pm$ 0.06 & 2.60 $\pm$ 0.57 &      ---   ---     &   0.79 $\pm$ 0.02   & 1.01 $\pm$ 0.05&  1.12 $\pm$ 0.25 &   ---  ---         &   1.35 $\pm$ 0.07 \\[0.07cm]  
\hline
         & & Region\,{\it C} &\scriptsize{$\langle A_V\rangle=5.93\pm 1.29$} & & & Region\,{\it D} &\scriptsize{$\langle A_V\rangle=3.99\pm 1.29$} & & \\[0.07cm]
\hline
1-0\,S(0)   &2.22  & 0.68 $\pm$ 0.22 & 1.23 $\pm$ 0.43 &  8.85 $\pm$ 2.12  & 1.52 $\pm$ 0.49  & 1.21 $\pm$ 0.23 & 1.82 $\pm$ 0.42 &  14.18 $\pm$ 2.02 &  4.87 $\pm$ 0.92   \\[0.07cm]
1-0\,S(1)   &2.12  & 2.71 $\pm$ 0.27 & 5.17 $\pm$ 0.90 & 10.48 $\pm$ 0.76  & 5.65 $\pm$ 0.57  & 4.99 $\pm$ 0.27 & 7.71 $\pm$ 1.16 &  15.43 $\pm$ 0.58 & 20.82 $\pm$ 1.12   \\[0.07cm]
1-0\,S(2)   &2.03  & 1.42 $\pm$ 0.36 & 2.82 $\pm$ 0.84 & 16.46 $\pm$ 3.16  & 2.66 $\pm$ 0.68  & 1.76 $\pm$ 0.28 & 2.80 $\pm$ 0.61 &  17.02 $\pm$ 1.97 &  6.56 $\pm$ 1.04   \\[0.07cm]
1-0\,S(3)   &1.96  & 3.07 $\pm$ 0.34 & 6.39 $\pm$ 1.24 & 13.58 $\pm$ 1.10  & 5.29 $\pm$ 0.58  & 4.99 $\pm$ 0.31 & 8.16 $\pm$ 1.39 &  16.80 $\pm$ 0.74 & 16.51 $\pm$ 1.02   \\[0.07cm]
2-1\,S(1)   &2.25  & 0.56 $\pm$ 0.25 & 1.01 $\pm$ 0.47 & 11.67 $\pm$ 3.96  & 1.32 $\pm$ 0.58  & 0.90 $\pm$ 0.23 & 1.34 $\pm$ 0.38 &  15.47 $\pm$ 2.98 &  3.75 $\pm$ 0.94   \\[0.07cm]
2-1\,S(3)   &2.07  & 0.28 $\pm$ 0.23 & 0.54 $\pm$ 0.45 &  8.54 $\pm$ 5.23  & 0.54 $\pm$ 0.44  & 0.50 $\pm$ 0.19 & 0.78 $\pm$ 0.33 &  11.38 $\pm$ 3.30 &  1.88 $\pm$ 0.73   \\[0.07cm]
Br$\gamma$  &2.16  & 6.17 $\pm$ 0.31 &11.55 $\pm$ 1.68 & 12.20 $\pm$ 0.43  &13.31 $\pm$ 0.68  & 5.21 $\pm$ 0.26 & 7.94 $\pm$ 1.15 &  14.12 $\pm$ 0.49 & 21.17 $\pm$ 1.06   \\[0.07cm]
Br$\delta$  &1.94  & 3.14 $\pm$ 0.35 & 6.58 $\pm$ 1.29 & 10.18 $\pm$ 0.82  & 5.31 $\pm$ 0.59  & 2.92 $\pm$ 0.29 & 4.81 $\pm$ 0.91 &  11.83 $\pm$ 0.83 &  9.28 $\pm$ 0.93   \\[0.07cm] 
$\rm [SiVI]$&1.96  & ---  ---        & ---  ---        &  --- ---          &---        ---    &    --- ---      &    --- ---      &    ---  ---       & --- ---            \\[0.07cm ]
$\rm [FeII]$&1.64  &11.00 $\pm$ 0.84 &28.78 $\pm$ 6.40 & 10.97 $\pm$ 0.60  & 6.64 $\pm$ 0.51  &27.40 $\pm$ 1.05 &52.33 $\pm$11.12 &  28.74 $\pm$ 0.79 & 40.94 $\pm$ 1.57   \\[0.07cm] 
HeI         &2.06  & 1.81 $\pm$ 0.29 & 3.56 $\pm$ 0.78 & 11.83 $\pm$ 1.42  & 3.44 $\pm$ 0.56  & 1.72 $\pm$ 0.24 & 2.71 $\pm$ 0.55 &  14.92 $\pm$ 1.55 &  6.35 $\pm$ 0.90   \\[0.07cm]
CO\tablenotemark{c}  &2.29  & 5.86 $\pm$ 0.31 & 6.39 $\pm$ 1.43 &  --- ---          & 5.93 $\pm$ 0.31  & 3.03 $\pm$ 0.21 & 3.31 $\pm$ 0.76 &    ---  ---       & 5.46 $\pm$ 0.38   \\[0.07cm] 
NaI\tablenotemark{c}    &2.21  & 1.66 $\pm$ 0.07 & 1.83 $\pm$ 0.41 &  --- ---          & 1.56 $\pm$ 0.07  & 0.93 $\pm$ 0.05 & 1.03 $\pm$ 0.23 &    ---  ---       &  1.64 $\pm$ 0.1    \\[0.07cm]
\hline
        &  & Region\,{\it E}  &\scriptsize{$\langle A_V\rangle=4.75\pm 1.29$} & & & Region\,{\it F} &\scriptsize{$\langle A_V\rangle=5.61\pm 1.34$} & &   \\[0.07cm]
\hline
1-0\,S(0)       &2.22  & 1.49 $\pm$ 0.29  &  2.41 $\pm$ 0.57 &  13.76 $\pm$ 2.06  & 3.59 $\pm$ 0.70  &   0.66 $\pm$ 0.16  &  1.16 $\pm$ 0.32  &   10.54 $\pm$ 1.83  &   3.78 $\pm$ 0.89 \\[0.07cm]
1-0\,S(1)       &2.12  & 7.68 $\pm$ 0.36  & 12.89 $\pm$ 1.91 &  18.47 $\pm$ 0.62  &19.04 $\pm$ 0.91  &   2.62 $\pm$ 0.20  &  4.84 $\pm$ 0.80  &   12.20 $\pm$ 0.66  &  13.86 $\pm$ 1.08 \\[0.07cm]
1-0\,S(2)       &2.03  & 3.20 $\pm$ 0.40  &  5.56 $\pm$ 1.09 &  21.30 $\pm$ 1.96  & 7.05 $\pm$ 0.88  &   1.10 $\pm$ 0.23  &  2.12 $\pm$ 0.55  &   14.43 $\pm$ 2.17  &   5.15 $\pm$ 1.06 \\[0.07cm]
1-0\,S(3)       &1.96  & 8.87 $\pm$ 0.42  & 15.96 $\pm$ 2.65 &  18.42 $\pm$ 0.63  &17.48 $\pm$ 0.83  &   2.80 $\pm$ 0.23  &  5.60 $\pm$ 1.04  &   13.31 $\pm$ 0.79  &  12.10 $\pm$ 1.01 \\[0.07cm]
2-1\,S(1)       &2.25  & 1.38 $\pm$ 0.45  &  2.22 $\pm$ 0.78 &  24.00 $\pm$ 6.61  & 3.48 $\pm$ 1.13  &   0.39 $\pm$ 0.13  &  0.69 $\pm$ 0.25  &    8.68 $\pm$ 2.18  &   2.37 $\pm$ 0.81 \\[0.07cm]
2-1\,S(3)       &2.07  & 0.71 $\pm$ 0.28  &  1.21 $\pm$ 0.51 &  13.89 $\pm$ 4.12  & 1.56 $\pm$ 0.62  &   0.29 $\pm$ 0.15  &  0.55 $\pm$ 0.29  &    8.83 $\pm$ 3.29  &   1.43 $\pm$ 0.73 \\[0.07cm]
Br$\gamma$      &2.16  & 4.07 $\pm$ 0.30  &  6.72 $\pm$ 1.04 &  14.23 $\pm$ 0.74  & 9.87 $\pm$ 0.72  &   1.86 $\pm$ 0.19  &  3.36 $\pm$ 0.59  &   13.52 $\pm$ 1.00  &  10.26 $\pm$ 1.06 \\[0.07cm]
Br$\delta$      &1.94  & 2.17 $\pm$ 0.32  &  3.92 $\pm$ 0.86 &  10.28 $\pm$ 1.12  & 4.09 $\pm$ 0.61  &   1.14 $\pm$ 0.22  &  2.30 $\pm$ 0.59  &   10.19 $\pm$ 1.41  &   4.81 $\pm$ 0.93 \\[0.07cm]
$\rm [SiVI]$    &1.96  &    ---  ---      &    ---  ---      &   ---   ---        & ---   ---        &   0.51 $\pm$ 0.20  &  1.02 $\pm$ 0.43  &   11.42 $\pm$ 3.23  &   2.11 $\pm$ 0.81 \\[0.07cm]
$\rm [FeII]$    &1.64  &57.34 $\pm$ 1.58  &123.91 $\pm$26.12 &  33.54 $\pm$ 0.68  &50.59 $\pm$ 1.39  &  10.50 $\pm$ 0.72  & 26.10 $\pm$ 5.96  &   18.03 $\pm$ 0.88  &  15.20 $\pm$ 1.04 \\[0.07cm]
HeI             &2.06  & 1.40 $\pm$ 0.30  &  2.41 $\pm$ 0.62 &  14.45 $\pm$ 2.28  & 3.06 $\pm$ 0.65  &   0.68 $\pm$ 0.18  &  1.28 $\pm$ 0.40  &   11.90 $\pm$ 2.41  &   3.23 $\pm$ 0.88 \\[0.07cm] 
CO\tablenotemark{c}      &2.29  & 6.46 $\pm$ 0.37  & 7.04 $\pm$ 1.58 &   ---   ---        &  7.00 $\pm$ 0.40  &   1.66 $\pm$ 0.14  &  1.82 $\pm$ 0.42  &   ---   ---         &  4.44 $\pm$ 0.37 \\[0.07cm]
NaI\tablenotemark{c}         &2.21  & 1.93 $\pm$ 0.09  &  2.13 $\pm$ 0.47 &   ---   ---        & 2.01 $\pm$ 0.09  &   0.60 $\pm$ 0.04  &  0.66 $\pm$ 0.15  &   ---   ---         &   1.51 $\pm$ 0.10 \\[0.07cm]
\hline
\enddata
\\[0.2cm]
\tablenotetext{a}{Total measured fluxes within aperture of $\rm 0.62^{\prime
    \prime}$ ($0.54^{\prime \prime}$ for [FeII] in {\it H}-band) line fluxes.}
\tablenotetext{b}{Galactic extinction has not been applied given the low value
    of $E(B-V) = 0.06$ (Schlegel et al.\ 1998). Total fluxes corrected by
    local extinction ($\langle A_V \rangle$, see Sec\,\ref{sss:subs232}).}
\tablenotetext{c}{Stellar absorption features. CO band corresponding to the
    2-1 transition at $\rm 2.29\,\mu m$. NaI doublet at $2.206$-$\rm
    2.208\,\mu m$.}

\end{deluxetable}
\clearpage
\end{landscape}

\end{document}